\documentclass[a4paper,11pt]{article}
\usepackage{amsmath}
\usepackage{a4wide}
\usepackage{multicol}
\usepackage{bm}
\usepackage{graphicx}
\usepackage{pstricks}
\usepackage{amsfonts}
\usepackage{mathrsfs}
\usepackage{cancel}
\usepackage{amssymb}
\usepackage{color}
\usepackage{upgreek}
\usepackage{caption}
\usepackage[numbers,super,sort&compress]{natbib}

\makeatletter
\DeclareRobustCommand\onlinecite{\@onlinecite}
\def\@onlinecite#1{\begingroup\let\@cite\NAT@citenum\citealp{#1}\endgroup}
\makeatother

\newcommand{\pfrac}[2]{\frac{\partial{#1}}{\partial{#2}}}
\newcommand{\ppfrac}[2]{\frac{\partial^2{#1}}{\partial{#2}^2}}
\newcommand{\pppfrac}[2]{\frac{\partial^3{#1}}{\partial{#2}^3}}

\newcommand{\pxyfrac}[3]{\frac{\partial^2{#1}}{\partial{#2}\partial{#3}}}
\newcommand{\pxxyfrac}[3]{\frac{\partial^3{#1}}{\partial{#2}^2\partial{#3}}}
\newcommand{\pxyyfrac}[3]{\frac{\partial^3{#1}}{\partial{#2}\partial{#3}^2}}
\newcommand{\refe}[1]{(\ref{#1})}

\newcommand{\td}[1]{\tilde{#1}}
\newcommand{\infinity}{\infty}
\newcommand{\Ca}{\operatorname{Ca}}
\newcommand{\Rey}{\operatorname{Re}}

\newcommand{\pilfrac}[2]{\partial_{#2} {#1}}
\newcommand{\ppilfrac}[2]{\partial^2_{#2} {#1}}

\newcommand{\bmN}{\bm{\nabla}}
\newcommand{\bmDot}{\bm{\cdot}}
\newcommand{\bmtau}{\bm{\tau}}
\newcommand{\tbU}{\vect{u}}

\newcommand{\tbN}{\vect{n}}
\newcommand{\ilfrac}[2]{{#1}/{#2}}
\newcommand{\lr}[1]{\left( {#1} \right)}
\newcommand{\lrsq}[1]{\left[ {#1} \right]}
\newcommand{\lrcur}[1]{\left\{ {#1} \right\}}
\newcommand{\lreval}[1]{\left. {#1} \right|}
\newcommand{\lrmod}[1]{\left| {#1} \right|}
\newcommand{\Cn}{\operatorname{Cn}}
\newcommand{\eg}{\textit{e.g.}}
\newcommand{\ie}{\textit{i.e.}}
\newcommand{\inin}[1]{\td{#1}}
\newcommand{\inint}[1]{\breve{#1}}
\newcommand{\upd}{\textrm{d}}
\newcommand{\tens}[1]{\textbf{#1}}
\newcommand{\vect}[1]{\textbf{#1}}
\newcommand{\etal}{\textit{et al.} }
\newcommand{\bfone}{\xi}
\newcommand{\bftwo}{\eta}
\newcommand{\bftwoin}{\zeta}

\newcommand{\tbR}{\bar{\textbf{r}}}
\newcommand{\fhs}{f_{\textrm{HS}}}
\newcommand{\nr}{\bar{\rho}(\tbR)}
\newcommand{\nrp}{\bar{\rho}(\tbR')}
\newcommand{\Vr}{V_p(\tbR)}
\newcommand{\dr}{\upd\tbR}
\newcommand{\drp}{\upd\tbR'}
\newcommand{\pattr}{\phi_{\textrm{attr}}}
\newcommand{\rmrp}{\lrmod{\tbR-\tbR'}}

\begin{document}

\title{The contact line behaviour of solid-liquid-gas diffuse-interface models}
\author{David N. Sibley$^1$, Andreas Nold$^1$, Nikos Savva$^{2,1}$, and Serafim Kalliadasis$^1$
\vspace{0.5cm}
\\
\small{$^1$ Department of Chemical Engineering, Imperial College London, London SW7 2AZ, UK}\\
\small{$^2$ School of Mathematics, Cardiff University, Senghennydd Road, Cardiff CF24 4AG, UK}
}
\date{Published: Phys. Fluids \textbf{25}, 092111 (2013); DOI: 10.1063/1.4821288}

%%%%%%%%%%%%%%%%%%%%%%%%%%%%%%%

\maketitle

\begin{abstract}
A solid-liquid-gas moving contact line is considered through a
diffuse-interface model with the classical boundary condition of no-slip at
the solid surface. Examination
of the asymptotic behaviour as the contact
line is approached shows
that the relaxation of the classical model of a
sharp liquid-gas interface, whilst retaining the no-slip condition, resolves
the stress and pressure singularities associated with the moving contact line
problem while the fluid velocity is well defined (not multi-valued). The
moving contact line behaviour is analysed for a general problem relevant for
any density dependent dynamic viscosity and volume viscosity, and for general
microscopic contact angle and double well free-energy forms. Away from the
contact line, analysis of the diffuse-interface model shows that the
Navier--Stokes equations and classical interfacial boundary conditions are
obtained at leading order in the sharp-interface limit, justifying the
creeping flow problem imposed in an intermediate region in the seminal work
of Seppecher [Int. J. Eng. Sci. \textbf{34}, 977--992 (1996)]. Corrections to
Seppecher's work are given, as an incorrect solution form was originally
used.
\end{abstract}

%%%%%%%%%%%%%%%%%
%
% INTRODUCTION
%
%%%%%%%%%%%%%%%%%

\section{Introduction}

A contact line occurs where solid, a liquid and gas (or another immiscible
liquid) meet. When the solid is in motion and the liquid-gas interface is
static, such as in coating processes, or when the solid is static and the
liquid-gas interface moves, such as when droplets spread under capillary
forces, the contact line is in motion, and differing merely through the
choice of reference frame. Moving contact lines are an important feature of a
vast range of both natural and technological processes, and as such modelling
and understanding of their behaviour is crucial to describe phenomena as
disparate as insects walking on water to oil recovery and inkjet printing.

Moving contact lines are also of considerable interest from the fundamental
point of view. The \emph{moving contact line problem} remains a persistent
long-standing problem in the field of fluid dynamics, despite its apparent
conceptual simplicity (see \eg~review articles by Dussan V.,\cite{Dussan79}
de Gennes,\cite{deGennesrev} Blake,\cite{blake2006physics} and Bonn
\etal\cite{BonnEggers}). The problem occurs as the static interface, modelled
classically as a sharp transition between liquid and gas, and the moving
solid surface, modelled with no-slip, cause an obvious multi-valued behaviour
in the velocity.\cite{DussanDavis,ShikhSingualarities06} This was highlighted
in the celebrated analysis of Huh and Scriven\cite{HuhScriv71} a number of
decades ago, encapsulating the moving contact line problem as the problem of
a \emph{non-integrable stress singularity} through analysing a planar liquid-
fluid interface meeting a solid, and showing the nonphysical prediction that
an infinite force would be required, if the classical model held, to submerge
a solid object.

In their concluding remarks, Huh and Scriven eloquently describe the failures
and omissions of the classical continuum mechanical model, many of which have
been fruitful areas of research in the subsequent years. The obvious culprit
for the unphysical singularities was first suggested to be the no-slip
condition at the wall. To resolve the problem some form of slip in the
contact line vicinity may be allowed, with Navier-slip, written down in the
early 19th Century by Navier,\cite{Navier} being proposed by Huh and
Scriven.\cite{HuhScriv71} The introduction of slip was not, however, proposed
by the authors as a physical effect, but rather that it would alleviate the
mathematical singularities with the boundary condition serving to parametrize
a number of microscale ingredients missing in the classical model in a
concise and amenable fashion. The range of microscale ingredients suggested
include the possibility of cavitation at the contact line, gradients in
viscosity and density---and appreciable compressibility effects,
non-Newtonian effects, the role of surfactants or heat accumulation, and
{long-range intermolecular forces, which are often
associated with the existence of a precursor film model ahead of the
macroscopic liquid-fluid front (a comparison between slip and precursor film
models was undertaken in Ref.~\onlinecite{SavvaPrecursorSlip}).}

The focus of this work is to consider a moving solid-liquid-gas contact line
but with the interface between liquid and gas to be diffuse.
Diffuse-interface models, reviewed \eg~by Anderson, McFadden, and
Wheeler,\cite{anderson_rev} have been around for a number of years, but have
been growing in popularity to consider the contact line since the crucial
works of Seppecher\cite{seppecher} for liquid-gas systems, and
Jacqmin\cite{jacqmin} for binary fluids. The seminal study of Seppecher, in
particular, will provide the foundation for the governing equations
considered here, however there are a number of important deficiencies
(further discussion of which are in Sec.~\ref{sec:seppecher}) and areas to
extend. Primarily, Seppecher's work is often referred to when suggesting that
diffuse-interface models resolve the moving contact line problem. Whilst the
work by Seppecher certainly contains some discussion of the asymptotics, the
analysis was somewhat incomplete, with asymptotic regions being probed
without careful justification and the crucial behaviour close to the contact
line only investigated numerically. A number of (unnecessary) constraints
were also imposed \eg~imposing a 90$^\circ$ contact angle and considering the
two fluids, \ie~the liquid and gas, to have equal viscosity.
Full numerical computations for the liquid-gas problem have been undertaken,
such as by Briant\cite{BriantYeomansEarly} and Briant, Wagner and
Yeomans,\cite{BriantYeomans1} using Lattice-Boltzmann methods. Computations
for diffuse-interface models for binary fluids {(\ie~for
solid-liquid-liquid contact line problems)}
have also been reported by a
number of authors; this is a different situation, however, often modelled as
two incompressible fluids, and including a coupled Cahn--Hilliard equation
{(with an extra nondimensional number, the mobility)}
to
account for the diffusion between the two fluid components and evolve this
effective order
parameter.\cite{jasnowvinals,KhatavkarJFM,DingSpeltJFM,YueZhouFeng}

Here, we examine analytically a diffuse-interface model without slip, a
precursor film, or any other ingredients. The classical model considers the
fluid-fluid interface to be a sharp surface of zero thickness where
quantities such as the fluid density are, in general, discontinuous. The
diffuse-interface approach relaxes this, in line with the physics of the
problem and in agreement with developments and applications in the field of
statistical mechanics of liquids {and in
molecular simulations, with quantities varying smoothly but rapidly (\eg~see
Refs.~\onlinecite{EvansReview,HendersonBook,Andreas1,yatsyshin1}), and
considers the interface to have a non-zero thickness, say $L_{\rm meso}$,
\ie~a mesoscopic or effective lengthscale (``mean field'') which is not known
a priori. In reality, this is of the order of the molecular scale, $L_{\rm
molecular}$, as also predicted by statistical mechanics and molecular
simulations, but diffuse-interface models do not involve any molecular
description. Nevertheless, $L_{\rm meso}/L_{\rm macro}\ll1$ giving the required
separation of scales and justifying the use of continuum equations for these models.
To recover the usual hydrodynamic equations and boundary conditions, one then takes the
sharp-interface limit, $L_{\rm meso}/L_{\rm macro} \rightarrow 0$.
In the context of moving contact line problems, the sharp-interface limit has been
considered carefully in analytical works with simple phase-field models,
and in numerical studies of phase-field, liquid-gas, and binary
fluid diffuse-interface models
\eg~Refs.~\onlinecite{AlikakosBatesChen,YueZhouFeng,YueFengPoF,christophe_marc}.}

{
Despite their success in describing several
fluid flow problems, diffuse-interface models are not free of
controversy. Recent debate in a special issue of the European Physical
Journal - Special Topics,\cite{epj_entire} has prompted questions,
principally from the founder of one of the more complex continuum mechanical
models applied to contact line motion---the interface formation model of
Shikhmurzaev,\cite{ShikhBook} in discussion articles,
Refs.~\onlinecite{epj1,epj5,epj6,epj8,epj11}. His views were countered by a
number of authors, in
Refs.~\onlinecite{epj4,epjpismen,epj3,epj7,pomeauphasefield}. The special
issue was concluded by Blake,\cite{epj12} who proposed a more collaborative
stance towards modelling the moving contact line, including understanding of
it from the viewpoint of diffuse-interface models. It is clear that
diffuse-interface models are of current interest and are increasingly being
successfully applied by many authors in a
wide variety of settings, thus analysis to further their theoretical
understanding is much needed.}

The fluid density $\bar\rho$ acts as an order parameter in this liquid-gas
diffuse-interface setting, such that in the sharp-interface limit (\ie~as the
thickness of the interface, $\epsilon$, tends to zero) the two bulk fluids
satisfy $\bar\rho=\rho_L$ and $\bar\rho=\rho_V=0$, being liquid and vapour
respectively, where we consider the behaviour of the system with vapour phase
of negligible density (as \eg~in Ref.~\onlinecite{PismenPomeau}). The
liquid-gas interface may then be defined as the location where
$\bar\rho=(\rho_L+\rho_V)/2$, although other choices such as defining it as
the location of the Gibbs dividing interface would be possible. In this
study, we also allow for general terms in the free energy and for the density
dependence of the viscosities, so that equal and disparate bulk viscosities
are encompassed. A simpler setting was considered in
Ref.~\onlinecite{myEPJE}, where both viscosity and free energy forms were
specified, to elucidate the alleviation of the singularities associated with
the classical modelling of contact line motion by the diffuse-interface
model, and to provide a basis upon which to compare this approach to {the aforementioned}
interface formation model of Shikhmurzaev.\cite{ShikhBook} These simpler
forms will be referred to where appropriate, throughout this article.

Our study will proceed by introducing the governing equations and geometry of
interest in Sec.~\ref{sec:probspec}, followed by nondimensionalisation and
the analysis of the equilibrium solution in Sec.~\ref{sec:nd}. The fluid
behaviour away from the contact line region will be examined using asymptotic
analysis in the sharp-interface limit in Sec.~\ref{sec:silim}, justifying the
intermediate solution used by Seppecher,\cite{seppecher} details and
corrections to this being given in Sec.~\ref{sec:seppecher}. The crucial
behaviour at the contact line is then analysed in Sec.~\ref{sec:cl}, where
the pressure and stress singularities associated with the moving contact line
problem are shown to be resolved in this diffuse-interface model
analytically, with discussion and conclusions in Sec.~\ref{sec:conclusions}.

\begin{figure}[t]
\centering
	\includegraphics{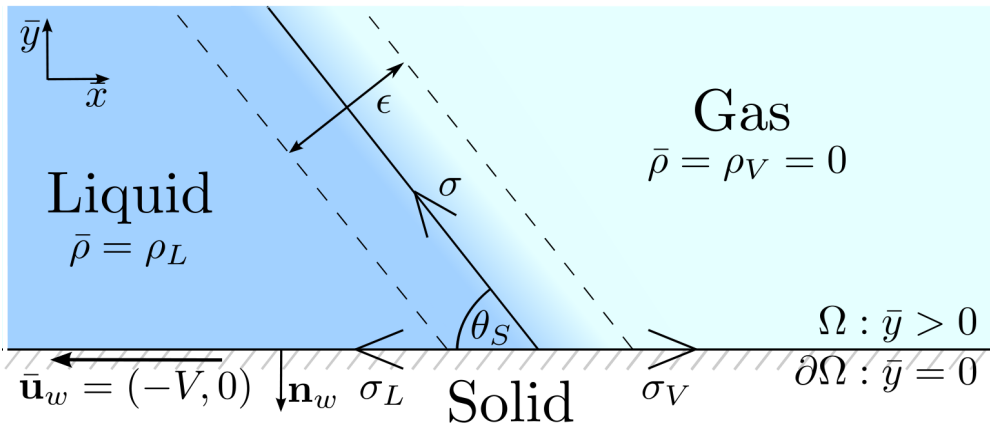}
	\caption{Sketch of the diffuse-interface model near a wall.}
\label{fig:diffinterface}
\end{figure}

\section{Problem specification}
\label{sec:probspec}

Consider the fluid domain $\Omega=\{ \bar{x}\in\mathbb{R}, 0<\bar{y} \}$
(bars are used to denote dimensional variables) and the solid surface
$\partial\Omega=\{ \bar{x}\in\mathbb{R}, \bar{y} = 0 \}$, where our frame of
reference is with the moving contact line, such that the driving force is the
wall velocity $\bar{\tbU}_w$, see Fig.~\ref{fig:diffinterface}. We assume
that the free energy of the system has contributions from an isothermal fluid
with a Helmholtz free energy functional and from the wall energy, given by
\begin{equation}
 \bar{\mathscr{F}} = \bar{\mathscr{F}}_m + \bar{\mathscr{F}}_w = \int_\Omega \bar{\mathcal{L}}  \upd\Omega + \int_{\partial\Omega} \bar{f}_w(\bar\rho)  \upd A, \qquad
 \bar{\mathcal{L}} = \bar\rho \bar{f}(\bar\rho) + \frac{1}{2}K|\bar\bmN\bar\rho|^2 -\bar{G}\bar\rho, \label{eq:origF}
\end{equation}
where $\bar{G}$ is the chemical potential, a Lagrange multiplier introduced to ensure mass conservation, $K$ is a gradient energy coefficient (assumed to be a constant for simplicity, as suggested in Ref.~\onlinecite{anderson_rev}), $\bar{f}_w(\bar\rho)$ is the wall energy, to be defined when considering the boundary conditions, and $\bar\rho \bar
f(\bar\rho)$ is a double well potential chosen to give the two
equilibrium states $\bar\rho=\{0,\rho_L\}$. In a similar way to a discussion in Ref.~\onlinecite{PismenPomeau}, $\epsilon = (K\rho_L/\breve{F})^{1/2}$ is the characteristic thickness of the interface, where $\breve{F}$ is a characteristic value of $\bar{f}(\bar\rho)$.

Forms such as in Eq.~\refe{eq:origF} for the free energy, and their
corresponding diffuse-interface approximations, have been adopted by numerous
authors for wetting problems such as Cahn,\cite{cahn77}
Seppecher,\cite{seppecher} Pismen and Pomeau,\cite{PismenPomeau} and
Pismen;\cite{Pismenmeso} see also the review by de Gennes.\cite{deGennesrev}
This form of free energy is simplified in that it uses a local approximation
to neglect non-local, integral, terms. The approximation effectively arises
through a Taylor series expansion of the density, and retaining up to and
including quadratic terms,\cite{Pismenmeso,antoniojfm} {further details
given in Appendix \ref{app:sqdf}.} The effect of such
non-local terms has been considered at equilibrium in studies such as in
Refs.~\onlinecite{Pismenmeso,antoniojfm}, where the long-range intermolecular
interactions are responsible for an algebraic decay of the density profile
away from the interface instead of the exponential one predicted by local
models (seen for the model considered here, in Eq.~\refe{eq:cak0}). In our
dynamic situation, we focus on the local approximation to elucidate the
contact line behaviour in a simplified, yet widely used and applicable
setting.

The density field augments the usual hydrodynamic equations through an additional stress tensor, termed the capillary or Korteweg tensor, defined as
\begin{align}
 \bar{\tens{T}} = \bar{\mathcal{L}}\tens{I} - \bar\bmN \bar\rho \otimes \pfrac{\bar{\mathcal{L}}}{(\bar\bmN \bar\rho)},
\end{align}
where $\tens{I}$ is the identity tensor, and which arises as a consequence of Noether's theorem.\cite{Noether,anderson_rev} On using the assumed form of $\bar{\mathcal{L}}$ from \refe{eq:origF} and coupling the capillary tensor with the usual viscous stress tensor $\bar\bmtau$ in the compressible Navier-Stokes equations, with total stress tensor $\bar{\tens{M}}$, the governing equations we thus consider are
\begin{gather}
 \pfrac{\bar\rho}{\bar{t}} + \bar\bmN\bmDot(\bar\rho\bar\tbU) = 0,\label{eq:origcty} \\
 \bar\rho \left( \pfrac{\bar\tbU}{\bar{t}} + \bar\tbU \bmDot\bar\bmN\bar\tbU \right) = \bar\bmN \bmDot \bar{\tens{M}},
\end{gather}
with
\begin{gather}
 \bar{\tens{M}} = \bar{\tens{T}} + \bar\bmtau,\label{eq:origM}\\
 \bar{\tens{T}} = \left( \bar\rho \bar f(\bar\rho) + \frac{K}{2}|\bar\bmN\bar\rho|^2 - \bar{G}\bar\rho \right)\tens{I} - K\bar\bmN\bar\rho\otimes\bar\bmN\bar\rho, \\
 \bar\bmtau = \bar\mu(\bar\rho)\left[(\bar\bmN\bar\tbU)+(\bar\bmN\bar\tbU)^\mathrm{T}\right]+ \bar\nu(\bar\rho)\left[ (\bar\bmN\bmDot\bar\tbU)\tens{I}\right], \label{eq:origtau}\\
 \bar{G} = -K\bar\nabla^2\bar\rho + \pfrac{}{\bar\rho}(\bar\rho \bar f(\bar\rho)),
\end{gather}
where $\bar\tbU$ is the fluid velocity, $\bar\mu(\bar\rho)$ and
$\bar\nu(\bar\rho)$ are the fluid viscosities (depending on the density), and
$\bar{t}$ is time. We have taken the viscous stress tensor $\bar\bmtau$ to be
Newtonian and the thermodynamic pressure is given by $\bar p=\bar\rho^2 \bar
f'(\bar\rho)$. The form for $\bar{G}$ arises from the Euler-Lagrange equation
corresponding to the free energy, and will be motivated by considering the
first variation of $\bar{\mathscr{F}}$ when discussing the wall boundary
conditions at the end of this section (the mathematical motivation of the
associated variational problem was discussed in the recent study in
Ref.~\onlinecite{Markus}).

For the analysis, we leave $\bar\mu(\bar\rho)$, $\bar\nu(\bar\rho)$ and
$\bar\rho \bar f(\bar\rho)$ general, but as mentioned in the introduction,
for ease of comprehension and clarity, we will include results for the
specific case considered in Ref.~\onlinecite{myEPJE}, and displayed
equations relating to this specific case will be labelled ``SC''. This
specific case used viscosity and density forms
\begin{align}
 \mbox{SC:} \qquad \bar\mu(\bar\rho) = \frac{\mu_L}{\rho_L} \bar\rho, \qquad \bar\nu(\bar\rho) = -\frac{2}{3}\bar\mu(\bar\rho)
 \qquad \bar f(\bar\rho) = \frac{K}{2\epsilon^2}\bar\rho\left( 1-\frac{\bar\rho}{\rho_L} \right)^2, \label{eq:dw_pot}
\end{align}
where $\bar\mu(\bar\rho)$ has the two equilibrium states at $\mu_L$ and $\mu_V = 0$.

On the solid surface $\partial\Omega$, with normal $\tbN_w$, we impose
\begin{equation}
 \bar{\tbU} = \bar{\tbU}_w, \qquad
 K \tbN_w \bmDot \bar\bmN \bar\rho + \bar f'_w(\bar\rho) = 0,
\end{equation}
with wall velocity $\bar{\tbU}_w=(-V,0)$ in Cartesian coordinates at $\bar{y}=0$. The first
condition is the classical no-slip, whilst the second arises through variational
arguments, discussed at the end of this section, and is termed the \textit{natural} (or \textit{wetting}) boundary
condition.\cite{YueZhouFeng04} {We note that this form of the natural boundary condition requires the wall free energy to instantaneously relax to equilibrium (as originally understood by Jacqmin~\cite{jacqmin}). In the conclusion, another form will be discussed where a finite timescale for the relaxation is posited, allowing for dynamic microscopic contact angle variation. Although the analysis does not change significantly, we proceed with the above condition to demonstrate that this form is sufficient to resolve the singularities associated with the moving contact line problem---without this extra degree of freedom.} The form for $\bar f_w$ should be chosen to satisfy Young's law at
the contact line, with solid-liquid, solid-gas and liquid-gas surface tensions $\bar
f_w(\rho_L)=\sigma_L$, $\bar f_w(0)=\sigma_V$ and $\sigma$, to be specified in \refe{eq:dimsurfaceT}, respectively,
and with contact angle $\theta_S$. A cubic is the lowest-order
polynomial required such that the wall free energy can be minimised for the
bulk densities and prevent a precursor film (or enrichment/depletion) forming away from the contact line, \ie~$\bar
f'_w(\rho_L)=\bar f'_w(0)=0$. Whilst cubic forms are used
for binary fluid problems, \eg~in Refs.~\onlinecite{jacqmin,YueZhouFeng}, this
is unlike the linear forms proposed in
the previous studies for liquid-gas problems\cite{seppecher,BriantYeomansEarly,BriantYeomans1,qianliqgas}, allowing us to
consider a diffuse-interface model without further physical effects from the microscale. These requirements determine
\begin{equation}
\bar f_w(\bar\rho) = \lrsq{{\rho_L^{-3}\sigma\cos\theta_S} ( 4\bar\rho^3 -6\bar\rho^2\rho_L + \rho_L^3 ) + \sigma_V+\sigma_L }/2,\label{eq:wallfdim}
\end{equation}
giving $\bar f'_w(\bar\rho) = -{6\sigma}\bar\rho\left( \rho_L - \bar\rho
\right)\cos\theta_S/{\rho_L^3}$, and $\bar f_w(0)-\bar f_w(\rho_L) =
\sigma\cos\theta_S$, with Young's law thus satisfied. It is noteworthy that
the natural boundary condition may be replaced by a constant density
condition if a precursor film/disjoining pressure model is to be considered,
\ie~$\bar{\rho}=\rho_a$ on $\partial\Omega$ (as used by Pismen and Pomeau\cite{PismenPomeau}), although as we shall demonstrate, this model is not
necessary to remove the singularities associated with the moving contact line.
Finally, for a one-dimensional density profile $\bar\rho(\bar{z})$ in
equilibrium the surface tension across the
interface is\cite{CH58}
\begin{equation}
 \sigma = K \int_{-\infinity}^\infinity\left( \frac{\upd\bar{\rho}}{\upd\bar{z}} \right)^2\, \upd\bar{z}
=\sqrt{2K}\int_{0}^{\rho_L}\sqrt{ \bar{\rho}[\bar{f}(\bar\rho)-\bar{G}_e] }\, \upd\bar{\rho},\label{eq:dimsurfaceT}
\end{equation}
which for the specific double well form in \refe{eq:dw_pot}, considered in Ref.~\onlinecite{myEPJE}, this gives
\begin{equation}
 \mbox{SC:} \qquad \sigma =  \frac{K\rho_L^2}{6\epsilon}\label{eq:dimsurfaceTs}.
\end{equation}
We note that equilibrium occurs in the governing equations when $\bar{\tbU}=0$ and when $\bar{G}=\bar{G}_e$, a constant which is chosen such that $\bar{\rho}\bar{f}(\bar\rho)$ predicts equally stable bulk fluids. In \refe{eq:dimsurfaceT}, $\bar{G}_e$ is unknown, but it may be assumed that $\bar{\rho}\bar{f}(\bar\rho)$ is chosen in such a way to require $\bar{G}_e=0$ (\ie~$[\bar\rho\bar{f}(\bar\rho)]|_{\bar\rho=\rho_L} = [\bar\rho\bar{f}(\bar\rho)]|_{\bar\rho=0}$), and in particular this is true of the form in \refe{eq:dw_pot}.

We now briefly describe how the natural boundary condition and the chemical
potential arise. Consider the first variation of the total free energy
$\bar{\mathscr{F}}$, i.e. consider
\begin{align}
 \lreval{\frac{\delta}{\delta\varepsilon}\lr{\bar{\mathscr{F}}[\bar{\rho}+\varepsilon{\bar{\rho}^*}]}}_{\varepsilon=0}=0,
\end{align}
where $\bar{\rho}^*$ is the increment of the density, so that
\begin{align}
 0=& \int_{\Omega}\lrcur{K(\bar\bmN\bar{\rho})\bmDot(\bar\bmN{\bar{\rho}^*}) + \lrsq{{\bar{\rho}^*} \bar{f}(\bar{\rho}) + \bar{\rho}{\bar{\rho}^*}\bar{f}'(\bar{\rho})}-\bar{G}{\bar{\rho}^*}} \upd\Omega
 + \int_{\partial\Omega}{\bar{\rho}^*}\bar{f}_w'(\bar{\rho}) \upd A,\\
\intertext{thus}
 0=& \int_{\Omega}\lrcur{K\lrsq{\bar\bmN\bmDot({\bar{\rho}^*}\bar\bmN\bar{\rho})-{\bar{\rho}^*}\bar\bmN\bmDot(\bar\bmN\bar{\rho})} +
 {\bar{\rho}^*} \bar{f}(\bar{\rho}) + \bar{\rho}{\bar{\rho}^*}\bar{f}'(\bar{\rho})-\bar{G}{\bar{\rho}^*}} \upd \Omega
 + \int_{\partial\Omega}{\bar{\rho}^*}\bar{f}_w'(\bar{\rho})  \upd A.\\
\intertext{Applying the divergence theorem, this becomes}
 0=& \int_{\Omega}-K{\bar{\rho}^*}\bar\bmN\bmDot(\bar\bmN\bar{\rho}) +
 {\bar{\rho}^*} \bar{f}(\bar{\rho}) + \bar{\rho}{\bar{\rho}^*}\bar{f}'(\bar{\rho})-\bar{G}{\bar{\rho}^*} \upd \Omega
 + \int_{\partial\Omega} K\tbN_w\bmDot({\bar{\rho}^*}\bar\bmN\bar{\rho}) + {\bar{\rho}^*} \bar{f}_w'(\bar{\rho}) \upd A, \nonumber\\
 =& \int_{\Omega}{\bar{\rho}^*}\lrsq{ \bar{f}(\bar{\rho})+\bar{\rho}\bar{f}'(\bar{\rho})-\bar{G} -K\bar\bmN\bmDot(\bar\bmN\bar{\rho}) }
 \upd \Omega
 + \int_{\partial\Omega} {\bar{\rho}^*}\lrsq{K\tbN_w\bmDot\bar\bmN\bar{\rho} + \bar{f}_w'(\bar{\rho})} \upd A, \nonumber\\
=& \int_{\Omega}{\bar{\rho}^*}\lrsq{ \pfrac{}{\bar{\rho}}[\bar{\rho}\bar{f}(\bar{\rho})]-\bar{G} -K\bar\nabla^2\bar{\rho} }
 \upd \Omega
 + \int_{\partial\Omega} {\bar{\rho}^*}\lrsq{K\tbN_w\bmDot\bar\bmN\bar{\rho} + \bar{f}_w'(\bar{\rho})} \upd A,
\end{align}
and as this is true for any increment ${\bar{\rho}^*}$, then the bulk chemical potential is specified by
\begin{align}
  \bar{G} = \pfrac{}{\bar{\rho}}[\bar{\rho}\bar{f}(\bar{\rho})] -K\bar\nabla^2\bar{\rho}, \qquad \mbox{in } \Omega,
\end{align}
and $\bar{\rho}$ must satisfy the natural boundary condition
\begin{align}
 K\tbN_w\bmDot\bar\bmN\bar{\rho} + \bar{f}_w'(\bar{\rho})=0, \qquad \mbox{on }\partial\Omega.
\end{align}
The normal $\tbN_w$ has arisen through application of the divergence theorem above, and thus must be defined as the outward unit normal to $\Omega$ on the boundary $\partial\Omega$. In Cartesian components and for our geometry $\tbN_w=(0,-1)$, and the natural boundary condition thus gives
\begin{align}
 -K\pfrac{\bar{\rho}}{\bar{y}} + \bar{f}_w'(\bar{\rho})=0, \qquad \mbox{on }\bar{y}=0.
\end{align}

\section{Nondimensionalization of the model and other representations}
\label{sec:nd}

We introduce the scalings
\begin{gather}
 \nonumber
 \bar{x} = X x, \quad \bar{y} = X y, \quad \bar\tbU = V \tbU, \quad \bar{t} = \frac{X}{V} t, \quad \bar{\rho} = \rho_L \rho, \quad
 \bar{p} = \frac{\mu_L V}{X} p, \\ \bar{G} = \frac{K\rho_L}{\epsilon X} G, \quad \bar{f} = \breve F f =
 \frac{K\rho_L}{\epsilon^2} f, \quad
 \bar{\tens{M}} = \frac{\mu_L V}{X} \tens{M}, \quad
 \bar{\tens{T}} = \frac{K\rho_L^2}{\epsilon X} \tens{T}, \quad \bar\bmtau = \frac{\mu_L V}{X} \bmtau,
\end{gather}
where $X$ is a typical length scale (which would be based on the macroscopic geometry, such as a channel width or droplet radius, for instance), $V$ is a typical velocity scale (we choose this based on the wall velocity in the $x$-direction), and we nondimensionalize $\bar\rho$ with the bulk liquid density. The other scalings arise from balances in the governing equations, and will contain the nondimensional parameters
\begin{equation}
 \Ca = \frac{\mu_L V}{\sigma}, \quad \Cn = \frac{\epsilon}{X}, \quad \Rey = \frac{\rho_L V X}{\mu_L},
 \quad \Ca_K = \frac{\mu_L V \epsilon}{K \rho_L^2},
\end{equation}
being the usual Capillary number, Cahn number and Reynolds number, and a
modified Capillary number based on the model parameter $K$, respectively. We
describe $\Ca_K$ as a modified Capillary number as for a given free-energy
form , the modified number and the usual one, are directly proportional
through the non-dimensional version of Eq.~\refe{eq:dimsurfaceT} (given later
in Eq.~\refe{eq:CakCaND}). The governing equations become
\begin{align}
\label{eq:dfge1}
 \pfrac{\rho}{t} + \bmN\bmDot(\rho\tbU) &= 0, \\
 \Rey \rho \left( \pfrac{\tbU}{t} + \tbU \bmDot\bmN\tbU \right) &= \bmN \bmDot \tens{M},\\
 \tens{M} &= \frac{1}{\Ca_K}\tens{T} + \bmtau, \\
 \tens{T} &= \left( \frac{1}{\Cn} \rho f(\rho) + \frac{\Cn}{2}|\bmN\rho|^2 - G\rho \right)\tens{I} -\Cn\bmN\rho\otimes\bmN\rho, \label{eq:dfge4}\\
 \bmtau &= \mu(\rho)\lrsq{(\bmN\tbU)+(\bmN\tbU)^\mathrm{T}} + \nu(\rho)(\bmN\bmDot\tbU)\tens{I}, \label{eq:dfge5}\\
 G &= -\Cn\nabla^2\rho + \frac{1}{\Cn}\pfrac{}{\rho}(\rho f(\rho)),\label{eq:dfge6}\\
 p&=(\Cn\Ca_K)^{-1}\rho^2f'(\rho),\label{eq:dfge7}
\end{align}
and we note the nondimensional general viscosity functions $\mu(\rho)$ and $\nu(\rho)$. As discussed, in contrast to the analysis in Ref.~\onlinecite{myEPJE} we leave the free energy component $f(\rho)$ general, and thus also the pressure, but note the specific case considered there had
\begin{align}
 \mbox{SC:} \qquad f(\rho) = \frac{1}{2}\rho(1-\rho)^2, \qquad \mbox{and thus} \qquad p = \frac{1}{2\Cn\Ca_K}\rho^2(1-3\rho)(1-\rho).\label{eq:ndfrho}
\end{align}
On the solid surface $\partial\Omega$, we have
\begin{align}
 \tbU = \tbU_w, \qquad
 \Cn\frac{\Ca}{\Ca_K} \tbN_w \bmDot \bmN \rho = 6\cos\theta_S(1-\rho)\rho ,\label{eq:dfgeend}
\end{align}
where $\tbU_w$ is the (now non-dimensional) wall velocity, satisfying $\tbU_w=(-1,0)$ in Cartesian components, and in equilibrium for the one dimensional variation of $\rho(z)$ across the interface (being the equivalent nondimensional expression of \refe{eq:dimsurfaceT})
\begin{align}
 \frac{\Ca_K}{\Cn\Ca} &= \int_{-\infinity}^\infinity\left( \frac{\upd\rho}{\upd z} \right)^2 \ \upd z =
 \int_{-\infinity}^\infinity \sqrt{\frac{2}{\Cn^2}\rho f} \frac{\upd\rho}{\upd z} \ \upd z =
 \frac{\sqrt{2}}{\Cn} \int_0^1 \sqrt{\rho f} \ {\upd\rho},\label{eq:nondimsurfaceT}
\end{align}
suggesting
\begin{align}
 {\Ca_K} = {\sqrt{2}}\Ca \int_0^1 \sqrt{\rho f} \ {\upd\rho},\label{eq:CakCaND}
\end{align}
which for the specific case gives
\begin{align}
 \mbox{SC:}\qquad{\Ca}= {6}{\Ca_K}, \qquad \Cn \tbN_w \bmDot \bmN \rho = \cos\theta_S(1-\rho)\rho,
\end{align}
so that in this particular instance there is no Capillary number, $\Ca$, as part of the governing equations or boundary conditions---the problem then depends only upon $\Cn$, $\Ca_K$ and $\Rey$. As mentioned, we will leave this general, however, and not choose a specific $f(\rho)$.

\subsection{The combined form of the model}

We can combine many of the equations into a single equation, {noting that a similar combined equation form has been used in Ref.~\onlinecite{PismenPomeau} when considering a diffuse-interface model with a precursor film in the lubrication approximation}. From \refe{eq:dfge5} we determine
\begin{align}
 \bmN \bmDot\bmtau &= \mu(\rho)\bmN^2\tbU + [\mu(\rho)+\nu(\rho)]\bmN(\bmN\bmDot\tbU)+
\lrsq{(\bmN\tbU)+(\bmN\tbU)^\mathrm{T}}\bmN(\mu(\rho)) + \bmN(\nu(\rho))(\bmN\bmDot\tbU)\tens{I} ,
\end{align}
and by using \refe{eq:dfge4} and \refe{eq:dfge6}
\begin{align}
 \bmN \bmDot \tens{T}&= \bmN \bmDot\left[
\left( \frac{1}{\Cn} \rho f(\rho) + \frac{\Cn}{2}|\bmN\rho|^2 +\Cn\rho\nabla^2\rho - \frac{\rho}{\Cn}\pfrac{}{\rho}(\rho f(\rho))  \right)\tens{I} -\Cn\bmN\rho\otimes\bmN\rho
 \right] \nonumber\\
&= \bmN \left( \frac{\Cn}{2}|\bmN\rho|^2 +\Cn\rho\nabla^2\rho - \Ca_K p  \right) -\Cn \bmN \bmDot(\bmN\rho\otimes\bmN\rho)
  \nonumber\\
&= \Cn\rho\bmN(\nabla^2\rho) - \Ca_K \bmN p,
\end{align}
so that for the creeping flow case, where inertial forces are negligible if compared to viscous forces, \ie~$\Rey\ll1$, we have
\begin{align}
 \bmN \bmDot \tens{M} &= \frac{1}{\Ca_K}\bmN \bmDot \tens{T} + \bmN \bmDot\bmtau \nonumber\\ &=
\frac{\Cn}{\Ca_K}\rho\bmN(\nabla^2\rho) - \bmN p + \mu(\rho)\bmN^2\tbU + [\mu(\rho)+\nu(\rho)]\bmN(\bmN\bmDot\tbU)\nonumber\\&\qquad\qquad\qquad\qquad\qquad
+\lrsq{(\bmN\tbU)+(\bmN\tbU)^\mathrm{T}}\bmN(\mu(\rho)) + \bmN(\nu(\rho))(\bmN\bmDot\tbU)\tens{I}
\nonumber\\
&=0,\label{eq:simpmom}
\end{align}
with $p=(\Cn\Ca_K)^{-1}\rho^2f'(\rho)$, the continuity equation \refe{eq:dfge1} coupled, and the boundary conditions on $\partial\Omega$ remaining as in Eq.~\refe{eq:dfgeend}.

\subsection{Equilibrium solution}

To draw comparisons to previous work, and to give a basis for comparison when the dynamic contact line behaviour is analysed, we consider the equilibrium behaviour of the system, corresponding to $\Ca_K\ll1$. It is clear from Eqs.~\refe{eq:dfge1}--\refe{eq:dfge7} that this corresponds to the capillary tensor dominating, and thus $\bmN\bmDot\tens{T}=0$. The equilibrium solution is thus governed by
\begin{equation}
{\Cn^2}\rho\bmN(\nabla^2\rho) = \bmN \lrsq{\rho^2f'(\rho)},
\end{equation}
subject to the natural boundary condition $-[\Cn\Ca/(6\Ca_K)]\pilfrac{\rho}{y} = \cos\theta_S(1-\rho)\rho$ at $y=0$, and the expected bulk behaviour $\rho\to\{0,1\}$ and $\nabla^2\rho\to0$ as $x\to\pm\infinity$. Considering the equation, we do not want the trivial solution, and so expansion of the gradient and simplifying leads to
\begin{align}
{\Cn^2}\bmN(\nabla^2\rho) = \lr{2f'(\rho)+\rho f''(\rho)}\bmN \rho.
\end{align}
This vector equation may be integrated once to obtain the scalar equation
\begin{align}
{\Cn^2}(\nabla^2\rho) = \rho f'(\rho) + f(\rho) = \pfrac{}{\rho}(\rho f(\rho)),
\end{align}
having removed arbitrary constants as we expect $\nabla^2\rho\to0$ and $\pilfrac{}{\rho}(\rho f(\rho))\to0$ (\ie~the double well potential is minimised) far away from the interface.

The solution subject to the above conditions for the specific function $f(\rho)$ in \refe{eq:ndfrho}, considered in Ref.~\onlinecite{myEPJE} is
\begin{equation}
 \mbox{SC:}\qquad \rho = \frac{1}{2}\lr{1-\tanh\lrsq{\frac{x\sin\theta_S+y\cos\theta_S}{2\Cn} }}, \label{eq:cak0}
\end{equation}
having also fixed the interface at $\rho=1/2$. This profile is planar and
at angle $\theta_S$. Similar hyperbolic tangent or exponential behaviours for equilibrium solutions have been seen in many diffuse-interface and phase-field models, such as in Refs.~\onlinecite{PismenPomeau,YueZhouFeng04,YueZhouFeng,QianWangShengJFM,QianWangShengGNBCfirst,jacqmin,CH58,BriantYeomans1,KhatavkarJFM,KassnerPaperSI}, it being a signature of diffuse-interface models with a local approximation.

As discussed in the introduction, the density variation for physical systems between liquid and gas occurs on a length scale which is much smaller than the macroscopic length scale, and hence $\Cn=\ilfrac{\epsilon}{X}\ll1$. The sharp-interface limit is given by the asymptotic behaviour
as $\Cn\to0$, and understanding the behaviour of the governing equations in this limit is of central importance when considering diffuse-interface models, as the classical continuum models (as given in \eg~Ref.~\onlinecite{batchelor}) should be recovered if correct predictions are to be found.

We will next undertake a careful asymptotic analysis of the outer solution
away from the interface, and of the interfacial region away from the wall
(using body fitted coordinates), to show that the expected sharp-interface
equations (the Navier--Stokes equations and the usual capillary surface
stress conditions) are indeed recovered.

\section{Sharp-interface limit away from the walls}
\label{sec:silim}

\subsection{The outer (bulk) regions}
\label{sec:siout}

We consider the governing equations \refe{eq:dfge1}--\refe{eq:dfge7}. In the outer region for the bulk fluids away from the interface, where variables are denoted with $^\pm$, we consider $\Cn\to0$ and take the expansions
\begin{align}
 \rho(t,x,y) &= \rho_0^\pm + \Cn \rho_1^\pm + \Cn^2 \rho_2^\pm + \cdots, &\nonumber
 G(t,x,y) &= {G}_0^\pm + \Cn {G}_1^\pm + \Cn^2 {G}_2^\pm + \cdots, \\
 \tbU(t,x,y) &= {\tbU}_0^\pm + \Cn {\tbU}_1^\pm + \Cn^2 {\tbU}_2^\pm + \cdots, &\nonumber
 p(t,x,y) &= {p}_{0}^\pm + \Cn {p}_{1}^\pm + \Cn^2 {p}_{2}^\pm + \cdots, \\
 \tens{M}(t,x,y) &= \tens{M}_0^\pm + \Cn \tens{M}_1^\pm + \Cn^2 \tens{M}_2^\pm + \cdots, &\nonumber
 \tens{T}(t,x,y) &= \tens{T}_0^\pm + \Cn \tens{T}_1^\pm + \Cn^2 \tens{T}_2^\pm + \cdots, \\
 \bmtau(t,x,y) &= \bmtau_0^\pm + \Cn \bmtau_1^\pm + \Cn^2 \bmtau_2^\pm + \cdots,
\end{align}
where for example $\rho_0^+$ is in the region of the liquid, see Fig.~\ref{fig:diffinterface}. At leading order, $O(\Cn^{-1})$, we find that
\begin{align}
 \rho_0^\pm f(\rho_0^\pm) = \pfrac{}{\rho}(\rho_0^\pm f(\rho_0^\pm)) = 0,
\end{align}
from the chemical potential equation, determining that the two free energy minima corresponding to the liquid and gas bulk values are the two solutions for $\rho_0^\pm$. This then implies
\begin{align}
 \rho_0^\pm f'(\rho_0^\pm) + f(\rho_0^\pm) = 0 \quad\Rightarrow\quad \rho_0^{\pm2} f'(\rho_0^\pm) + \rho_0^\pm f(\rho_0^\pm) = 0 \quad\Rightarrow\quad
 \rho_0^{\pm2} f'(\rho_0^\pm) = 0,
\end{align}
being consistent with zero pressure at this order. The $O(1)$ equations then give
\begin{align}
 \rho_0^\pm\bmN\bmDot \tbU_0^\pm &= 0, \\
 \Rey \rho_0^\pm \left( \pfrac{\tbU_0^\pm}{t} + \tbU_0^\pm \bmDot\bmN\tbU_0^\pm \right) &= \bmN \bmDot \tens{M}_0^\pm,\\
 \tens{M}_0^\pm &= \frac{1}{\Ca_K}\tens{T}_0^\pm + \bmtau_0^\pm, \\
 \tens{T}_0^\pm &= \left( \rho_1^\pm (\rho_0^\pm f'(\rho_0^\pm)+f(\rho_0^\pm)) - G_0^\pm\rho_0^\pm \right)\tens{I}
= - G_0^\pm\rho_0^\pm \tens{I}
, \\
 \bmtau_0^\pm &= \mu(\rho_0^\pm)\lrsq{(\bmN\tbU_0^\pm)+(\bmN\tbU_0^\pm)^\mathrm{T}} + \nu(\rho_0^\pm) (\bmN\bmDot\tbU_0^\pm)\tens{I}, \\
 G_0^\pm &= \rho_1^\pm (\rho_0^\pm f''(\rho_0^\pm) + 2 f'(\rho_0^\pm)), \\
 p_{0}^\pm &= \frac{1}{\Ca_K}\rho_0^\pm\rho_1^\pm(\rho_0^\pm f''(\rho_0^\pm)+2f'(\rho_0^\pm)),
\end{align}
which simplify to
\begin{align}
 \rho_0^\pm\bmN\bmDot \tbU_0^\pm = 0, \qquad
 \Rey \rho_0^\pm \left( \pfrac{\tbU_0^\pm}{t} + \tbU_0^\pm \bmDot\bmN\tbU_0^\pm \right) = \bmN \bmDot \tens{M}_0^\pm,\qquad
 \tens{M}_0^\pm = -p_{0}^\pm\tens{I} + \bmtau_0^\pm,
\end{align}
since
\begin{align}
 \tens{T}_0^\pm = -G_0^\pm\rho_0^\pm \tens{I}, \qquad \mbox{and} \qquad G_0^\pm = {\Ca_K} \lr{\frac{p_{0}^\pm}{\rho_0^\pm}} .
\end{align}
This is an important result, as we see that at leading order the governing equations are the expected incompressible Navier--Stokes equations. The equations at the next order, $O(\Cn)$, are
\begin{align}
 \pfrac{\rho_1^\pm}{t} + \bmN\bmDot(\rho_1^\pm\tbU_0^\pm) + \rho_0^\pm\bmN\bmDot\tbU_1^\pm &= 0, \\
 \Rey \left[\rho_1^\pm \left( \pfrac{\tbU_0^\pm}{t} + \tbU_0^\pm \bmDot\bmN\tbU_0^\pm \right) + \rho_0^\pm \left( \pfrac{\tbU_1^\pm}{t} + \tbU_1^\pm \bmDot\bmN\tbU_0^\pm + \tbU_0^\pm \bmDot\bmN\tbU_1^\pm\right)\right] &= \bmN \bmDot \tens{M}_1^\pm,
\end{align}
where
\begin{align}
 \tens{M}_1^\pm &= \frac{1}{\Ca_K}\tens{T}_1^\pm + \bmtau_1^\pm, \\
 \tens{T}_1^\pm &= \left( \rho_2^\pm\lrsq{\rho_0^\pm f'(\rho_0^\pm)+f(\rho_0^\pm)} + \rho_1^{\pm2}\lrsq{f'(\rho_0^\pm)+\frac{1}{2}\rho_0^\pm f''(\rho_0^\pm)} - G_1^\pm\rho_0^\pm - G_0^\pm\rho_1^\pm\right)\tens{I} ,
\\
 \bmtau_1^\pm &= \mu(\rho_0^\pm)\lrsq{(\bmN\tbU_1^\pm)+(\bmN\tbU_1^\pm)^\mathrm{T}} + \nu(\rho_0^\pm)(\bmN\bmDot\tbU_1^\pm)\tens{I}
+ \rho_1^\pm\mu'(\rho_0^\pm)\lrsq{(\bmN\tbU_0^\pm)+(\bmN\tbU_0^\pm)^\mathrm{T}} + \rho_1^\pm\nu'(\rho_0^\pm)(\bmN\bmDot\tbU_0^\pm)\tens{I}
, \\
 G_1^\pm &= \frac{1}{2}\rho_1^{\pm2}\lrsq{3f''(\rho_0^\pm)+\rho_0^\pm f'''(\rho_0^\pm)} + \rho_2^\pm\lrsq{2f'(\rho_0^\pm)+\rho_0^\pm f''(\rho_0^\pm)}, \\
 p_{1}^\pm &= \frac{1}{\Ca_K}\lrcur{
\rho_0^\pm\rho_2^\pm\lrsq{2f'(\rho_0^\pm)+\rho_0^\pm f''(\rho_0^\pm)} + \rho_1^{\pm2}\lrsq{\frac{1}{2}\rho_0^{\pm2}f'''(\rho_0^\pm) + f'(\rho_0^\pm)+2\rho_0^\pm f''(\rho_0^\pm)}
},
\end{align}
being linear in $\rho_2^\pm$. The chemical potential and pressure terms may be used to determine
\begin{align}
 \tens{T}_1^\pm = -\Ca_K p_1^\pm\tens{I} ,
\end{align}
and thus a linearised form of the (compressible) Navier-Stokes equations is recovered.

\subsection{The interface region}
\label{sec:siin}

To consider the region near to the interface, the equations are transformed into body fitted coordinates $(\bfone,\bftwo)$. $\bfone$ measures distance along the interface (like an arc length), and $\bftwo$ measures the distance in the normal direction from a general point $(x,y)\in\Omega$ to the interface. For details see Appendix \ref{app:bfcoordinates}.

We take the stretched coordinate $\bftwoin = \bftwo/\Cn$ to examine the
region close to the interface. With this, we now have the interface
satisfying $\bftwoin=\bftwo=0$, and we note that $\bftwo>0$ in the liquid
(and $\bftwo<0$ in the gas) with $\vect{n}$ then pointing from gas to liquid.
We take the limit $\Cn\to0$ through the expansions
\begin{align}
 \rho &= \inint{\rho}_0 + \Cn \inint\rho_1 + \Cn^2 \inint\rho_2 + \cdots, &
 G &= \inint{G}_0 + \Cn \inint{G}_1 + \Cn^2 \inint{G}_2 + \cdots, \nonumber\\
 \tbU &= \inint{\tbU}_0 + \Cn \inint{\tbU}_1 + \Cn^2 \inint{\tbU}_2 + \cdots, &
 \bmtau &= \frac{1}{\Cn}\left[\inint{\bmtau}_0 + \Cn \inint{\bmtau}_1 + \Cn^2 \inint{\bmtau}_2 + \cdots\right], \nonumber\\
 \tens{T} &= \frac{1}{\Cn}\left[\inint{\tens{T}}_0 + \Cn \inint{\tens{T}}_1 + \Cn^2 \inint{\tens{T}}_2 + \cdots\right], &
 \tens{M} &= \frac{1}{\Cn}\left[\inint{\tens{M}}_0 + \Cn \inint{\tens{M}}_1 + \Cn^2 \inint{\tens{M}}_2 + \cdots\right], \nonumber\\
 p &= \frac{1}{\Cn}\left[\inint{p}_{0} + \Cn \inint{p}_{1} + \Cn^2 \inint{p}_{2} + \cdots\right]. \label{eq:inner_exps}
\end{align}
Matching between interfacial variables, denoted $\inint{\phi}$ for an arbitrary function $\phi$, and outer variables (denoted $\phi^\pm$) is next considered. For this arbitrary function, we require
$\phi^\pm(\bfone,\bftwo) = \phi^\pm(\bfone,\Cn\bftwoin) \sim \inint{\phi}(\bfone,\bftwoin)$ for $\bftwoin\to\infinity$, $\Cn\to0$, $\Cn\bftwoin\to0$. Thus
\begin{align}
 \phi^\pm(\bfone,\Cn\bftwoin) &= \phi^\pm_0(\bfone,0) + \Cn\left( \bftwoin\pfrac{\phi^\pm_0}{\bftwoin}(\bfone,0) + \phi^\pm_1(\bfone,0) \right) + \cdots, \nonumber\\
 \inint{\phi}(\bfone,\bftwoin) &= \inint{\phi}_0(\bfone,\bftwoin) + \Cn\inint{\phi}_1(\bfone,\bftwoin) + \cdots,
\end{align}
which gives the first two matching conditions at $O(1)$ and $O(\Cn)$ as
\begin{align}
 \lim_{\bftwoin\to\pm\infinity} \inint{\phi}_0(\bfone,\bftwoin) &= \phi_0^\pm(\bfone,0),\\
 \lim_{\bftwoin\to\pm\infinity} \inint{\phi}_1(\bfone,\bftwoin) &= \phi_1^\pm(\bfone,0) + \bftwoin\pfrac{\phi_0^\pm}{\bftwoin}(\bfone,0),
\end{align}
together implying
\begin{equation}
 \lim_{\bftwoin\to\pm\infinity} \pfrac{\inint{\phi_1}}{\bftwoin}(\bfone,\bftwoin) = \pfrac{\phi_0^\pm}{\bftwoin}(\bfone,0).
\end{equation}
Using the details from Appendix \ref{app:bfcoordinates}, we now go through each of the equations in turn and match to the outer region to determine the effective interfacial conditions in the sharp-interface limit.

\subsubsection{Continuity equation}
The continuity equation in body fitted coordinates and in the interfacial region is
\begin{align}
 \pfrac{{\rho}}{t} + \frac{u^s_t}{(1-\kappa\Cn\bftwoin)^2}\pfrac{{\rho}}{\bfone} +
 \frac{u^s_n}{\Cn}\pfrac{{\rho}}{\bftwoin} + \frac{1}{\Cn}\pfrac{({\rho} v)}{\bftwoin} -
 \frac{\kappa {\rho} v}{1-\kappa\Cn\bftwoin} + \pfrac{({\rho} {u})}{\bfone} =
 \Cn\frac{{\rho}{u}\bftwoin}{1-\kappa\Cn\bftwoin}\pfrac{\kappa}{\bfone} ,
\end{align}
where $u^s_t$, $u^s_n$ are the tangential and normal velocities of the interface respectively, and assumed to be $O(1)$. $\kappa$ is the interfacial curvature, and $(u,v)$ are the velocity components in the $(\bfone,\bftwoin)$ directions.

Considering expansions \refe{eq:inner_exps}, then the leading order of the continuity equation satisfies
\begin{align}
{u^s_n}\pfrac{\inint{\rho_0}}{\bftwoin} + \pfrac{(\inint{\rho}_0 \inint{v}_0)}{\bftwoin}=0.
\end{align}
Assuming that the normal velocity of the interface is independent of the coordinate normal to the interface $\bftwoin$, then the leading order equation suggests
\begin{align}
 \left[\inint{\rho_0}({u^s_n} + \inint{v}_0)\right]_{-\infinity}^\infinity = 0, \quad \mbox{which implies} \quad {v}^{+}_0(0) = -{u^s_n},
\end{align}
and being equivalent to
\begin{align}
 \pfrac{d}{t} + \tbU_0^\pm\bmDot\bmN d = 0,
\end{align}
the expected kinematic boundary condition for the sharp interface\cite{batchelor} defined by $d(\vect{x},t)=0$, where $d$ is the signed distance to the interface, equivalent to  $\bftwo$.

\subsubsection{Chemical potential equation}

We now consider the chemical potential equation in the interfacial region
\begin{align}
 \Cn G = \pfrac{}{\rho}(\rho f(\rho))-\Cn^2\left[ \frac{\ppilfrac{\rho}{\bfone}}{(1-\kappa\Cn\bftwoin)^2} + \frac{\Cn\bftwoin(\pilfrac{\kappa}{\bfone})}{(1-\kappa\Cn\bftwoin)^3}\pfrac{\rho}{\bfone}
 + \frac{\ppilfrac{\rho}{\bftwoin}}{\Cn^2} - \frac{1}{\Cn}\frac{\kappa(\pilfrac{\rho}{\bftwoin})}{1-\kappa\Cn\bftwoin} \right],
\end{align}
so that using the expansions \refe{eq:inner_exps}, we have at leading order
\begin{align}
 -\ppfrac{\inint{\rho}_0}{\bftwoin} + \lreval{\pfrac{}{\rho}(\rho f(\rho))}_{\rho=\inint{\rho}_0} = 0,
\end{align}
which may be solved by noting that $\pilfrac{}{\inint\rho_0}[ (\pilfrac{\inint\rho_0}{\bftwoin})^2 ] = 2\ppilfrac{\inint\rho_0}{\bftwoin}$ so that the equation becomes
\begin{align}
\label{eq:1storder:lorho}
 -\frac{1}{2}\left(\pfrac{\inint{\rho}_0}{\bftwoin}\right)^2 + \inint{\rho}_0 f(\inint{\rho}_0) = A = 0,
\end{align}
where the arbitrary constant of integration has been removed since
\begin{align}
 \lim_{\bftwoin\to\pm\infinity}(\inint{\rho}_0 f(\inint{\rho}_0))\to0, \qquad \mbox{and} \qquad
 \lim_{\bftwoin\to\pm\infinity}\pilfrac{\inint{\rho}_0}{\bftwoin}\to0.
\end{align}
In the specific case considered in Ref.~\onlinecite{myEPJE}, this may be solved to find the solution corresponding to $\lim_{\bftwoin\to\infinity}\inint{\rho}_0\to1$ and $\lim_{\bftwoin\to-\infinity}\inint{\rho}_0\to0$ of
\begin{align}
 \mbox{SC:} \qquad \inint{\rho}_0 = \frac{1}{2}\left[ 1+\tanh\left( \frac{\bftwoin}{2} \right) \right].
\end{align}
The following order terms suggest
\begin{align}
 \inint{G}_0 =  -\ppfrac{\inint{\rho}_1}{\bftwoin} + \kappa\pfrac{\inint{\rho}_0}{\bftwoin}
 + \inint{\rho_1}\lr{2f'(\inint\rho_0)+\inint\rho_0f''(\inint\rho_0)}. \label{eq:CPCn}
\end{align}
For a binary fluid model an analogue of this equation is critical, and a solvability condition arising from the application of the Fredholm alternative\cite{fredholm} is required to relate the chemical potential to the interface curvature. As we will see, in this liquid-gas setting the chemical potential enters into the next to leading order terms in the momentum, but a specific form is not required to obtain the interfacial conditions through matching to the outer regions in the sharp-interface limit.

\subsubsection{Stress tensors and pressure}

Considering the diffusion tensor $\tens{T}$ in the interfacial region, and using expansions \refe{eq:inner_exps}, gives
\begin{align}
\inint{\tens{T}}_0 = \left( \begin{array}{cc} 2\inint{\rho}_0f(\inint{\rho}_0) &0\\0&0 \end{array} \right),
\end{align}
after simplification using \refe{eq:1storder:lorho}. The components of the
next order, $\inint{\tens{T}}_1$ simplify to
\begin{align}
 (\inint{\tens{T}}_1)_{11} &= \inint\rho_1(\inint{\rho}_0f'(\inint{\rho}_0)+f(\inint{\rho}_0)) +
 \pfrac{\inint\rho_0}{\bftwoin}\pfrac{\inint\rho_1}{\bftwoin}-\inint{G}_0\inint{\rho}_0 + 4 \kappa \bftwoin  \inint{\rho}_0f(\inint{\rho}_0)
 , \nonumber\\
(\inint{\tens{T}}_1)_{12} &= (\inint{\tens{T}}_1)_{21} = -\pfrac{\inint\rho_0}{\bftwoin}\pfrac{\inint\rho_0}{\bfone},\nonumber\\
(\inint{\tens{T}}_1)_{22} &= -\pfrac{\inint\rho_0}{\bftwoin}\pfrac{\inint\rho_1}{\bftwoin}
+\inint\rho_1(\inint{\rho}_0f'(\inint{\rho}_0)+f(\inint{\rho}_0)) -\inint{G}_0\inint{\rho}_0.
\end{align}
Similarly, the viscous stress tensor $\inint\bmtau$ in the interfacial region gives
\begin{align}
\inint{\bmtau}_0 &= \left(\begin{array}{cc} \nu(\inint{\rho}_0)\pilfrac{v_0}{\bftwoin} & \mu(\inint{\rho}_0)\pilfrac{{u}_0}{\bftwoin} \\ \mu(\inint{\rho}_0)\pilfrac{{u}_0}{\bftwoin} & [2\mu(\inint{\rho}_0)+\nu(\inint{\rho}_0)]\pilfrac{v_0}{\bftwoin} \end{array}\right), \nonumber\\
(\inint{\bmtau}_1)_{11} &=
 [2\mu(\inint{\rho}_0)+\nu(\inint{\rho}_0)]\left(\pfrac{{u}_0}{\bfone}-\kappa v_0 \right)
 +\nu(\inint{\rho}_0) \left( \pfrac{v_1}{\bftwoin} + 2\kappa\bftwoin\pfrac{v_0}{\bftwoin}\right) + \inint{\rho}_1\nu'(\inint{\rho}_0)\lrsq{\pfrac{v_0}{\bftwoin}}, \nonumber\\
(\inint{\bmtau}_1)_{12} &= (\inint{\bmtau}_1)_{21} = \mu(\inint{\rho}_0)\lrsq{\pfrac{{u}_1}{\bftwoin} + \pfrac{v_0}{\bfone}} +
\inint{\rho}_1\mu'(\inint{\rho}_0)\lrsq{\pfrac{{u}_0}{\bftwoin}},
\nonumber\\
(\inint{\bmtau}_1)_{22} &= [2\mu(\inint{\rho}_0)+\nu(\inint{\rho}_0)]\pfrac{v_1}{\bftwoin} + \nu(\inint{\rho}_0)\left(\pfrac{{u}_0}{\bfone} - \kappa v_0\right) + \inint{\rho}_1[2\mu'(\inint{\rho}_0) + \nu'(\inint{\rho}_0) ]\lrsq{\pfrac{v_0}{\bftwoin}},
\end{align}
and the thermodynamic pressure (from \refe{eq:dfge7}), using expansions \refe{eq:inner_exps}, gives
\begin{align}
 \inint{p}_{0}=\frac{1}{\Ca_K}\inint\rho_0^2f'(\inint\rho_0), \qquad
 \inint{p}_{1}=\frac{1}{\Ca_K}\inint\rho_0\inint\rho_1(\inint\rho_0f''(\inint\rho_0)+2f'(\inint\rho_0)).
\end{align}
Matching suggests $\lim_{\bftwoin\to\pm\infinity}\inint{p}_{0} = 0$, and $\lim_{\bftwoin\to\pm\infinity}\inint{p}_{1} = p_{0}^\pm$.

\subsubsection{Momentum equation}

Considering the momentum equation in body fitted coordinates
\begin{align}
 \Rey \rho \left( \pfrac{\tbU}{t} + (\tbU^s + \tbU)\bmDot\bmN\tbU  \right) &=
\left( \begin{array}{c} \pilfrac{[(\tens{M})_{11}]}{\bfone}+\pilfrac{[(\tens{M})_{12}]}{\bftwo} + 2\Gamma_{11}^1(\tens{M})_{11} +
 3\Gamma_{12}^1(\tens{M})_{12} \\ \pilfrac{[(\tens{M})_{12}]}{\bfone}+\pilfrac{[(\tens{M})_{22}]}{\bftwo} + \Gamma_{11}^1(\tens{M})_{12} +
 \Gamma_{12}^1(\tens{M})_{22} + \Gamma_{11}^2(\tens{M})_{11} \end{array} \right),
\end{align}
where the Christoffel symbols are given in \refe{eq:christoffel}, and $\bmN\tbU$ in \refe{eq:nabu}. At leading order we have
\begin{align}
 \pfrac{}{\bftwoin}[(\inint{\tens{M}}_0)_{12}] = \pfrac{}{\bftwoin}[(\inint{\tens{M}}_0)_{22}]=0,
\end{align}
and integrating once suggests that these leading-order stress components are independent of $\bftwoin$. Applying the matching conditions we find that $\lim_{\bftwoin\to\pm\infinity}\inint{\tens{M}}_0=0$, and thus $(\inint{\tens{M}}_0)_{12}=(\inint{\tens{M}}_0)_{22}=0$, where
\begin{align}
 (\inint{\tens{M}}_0)_{11} = \frac{2}{\Ca_K}\inint{\rho}_0f(\inint{\rho}_0) + \nu(\inint{\rho}_0)\pfrac{\inint{v}_0}{\bftwoin}, \qquad
 (\inint{\tens{M}}_0)_{12} = \mu(\inint{\rho}_0)\pfrac{\inint{u}_0}{\bftwoin}, \qquad
 (\inint{\tens{M}}_0)_{22} = [2\mu(\inint{\rho}_0)+\nu(\inint{\rho}_0)]\pfrac{\inint{v}_0}{\bftwoin} ,
\end{align}
giving the result that $\inint{\tbU}_0$ is independent of $\bftwoin$, and thus giving the interface condition $\inint{\tbU}_0^+ = \inint{\tbU}_0^-$. The next order in the momentum equation then suggests
\begin{align}
  \Rey \inint{\rho}_0 \lrsq{ ({\tbU}^s + \inint{\tbU}_0)\bmDot \left( \begin{array}{cc} 0 & \pilfrac{\inint{u}_0}{\bftwoin} \\
 0 & \pilfrac{\inint{v}_0}{\bftwoin} \end{array} \right) } &=
\left( \begin{array}{c} \pilfrac{[(\inint{\tens{M}}_0)_{11}]}{\bfone}+\pilfrac{[(\inint{\tens{M}}_1)_{12}]}{\bftwoin} - 3\kappa (\inint{\tens{M}}_0)_{12} \\ \pilfrac{[(\inint{\tens{M}}_0)_{12}]}{\bfone}+\pilfrac{[(\inint{\tens{M}}_1)_{22}]}{\bftwoin} + \kappa[(\inint{\tens{M}}_0)_{11}-(\inint{\tens{M}}_0)_{22}] \end{array} \right),
\end{align}
and on using results found from the leading order solution, these reduce to
\begin{align}
 \pfrac{}{\bfone}[(\inint{\tens{M}}_0)_{11}]+\pfrac{}{\bftwoin}[(\inint{\tens{M}}_1)_{12}] = 0, \qquad
 \pfrac{}{\bftwoin}[(\inint{\tens{M}}_1)_{22}] + \kappa(\inint{\tens{M}}_0)_{11} = 0,
\end{align}
suggesting
\begin{align}
 \pfrac{}{\bftwoin}[(\inint{\tens{M}}_1)_{12}] = -\pfrac{}{\bfone}[(\inint{\tens{M}}_0)_{11}] =
 -\pfrac{}{\bfone}\lrsq{
2\frac{\inint{\rho}_0f(\inint{\rho}_0)}{\Ca_K} + \nu(\inint{\rho}_0)\pfrac{\inint{v}_0}{\bftwoin} }=
 -\pfrac{}{\bfone}\lrsq{\frac{1}{\Ca_K}\left(\pfrac{\inint{\rho}_0}{\bftwoin}\right)^2},
\end{align}
and
\begin{align}
 \pfrac{}{\bftwoin}[(\inint{\tens{M}}_1)_{22}] = -\kappa(\inint{\tens{M}}_0)_{11} = -\frac{\kappa}{\Ca_K}\left(\pfrac{\inint{\rho}_0}{\bftwoin}\right)^2.
\end{align}
Integrating these equations for the next order momentum we find
\begin{align}
 \left[ (\inint{\tens{M}}_1)_{12} \right]_{-\infinity}^\infinity = 0,
\end{align}
using
\begin{align}
 \int_{-\infinity}^\infinity \left(\pfrac{\inint{\rho}_0}{\bftwoin}\right)^2 \upd\bftwoin = \frac{\Ca_K}{\Ca},
\end{align}
being an interfacial region analogue of \refe{eq:nondimsurfaceT}, and
\begin{align}
 \left[ (\inint{\tens{M}}_1)_{22} \right]_{-\infinity}^\infinity = -\frac{\kappa}{\Ca},
\end{align}
and matching to the outer region gives the interface conditions
\begin{align}
 \tbN\bmDot\left[ \bmtau_0^+ - \bmtau_0^- \right]\bmDot(\tens{I}-\tbN\otimes\tbN)=0,\qquad
 \tbN\bmDot\left[\left(-p_{0}^+ + p_{0}^-\right)\tens{I} + \bmtau_0^+ - \bmtau_0^- \right]\bmDot\tbN = -\frac{\kappa}{\Ca}.
\end{align}
We have thus found that the classical equations and interfacial boundary
conditions are obtained at leading order in the sharp-interface limit,
justifying the assumption that they may be used in an intermediate region
away from the contact line, as done by Seppecher.\cite{seppecher} Whilst
this assumption has been justified, there are some important points to note,
which are discussed next.

\section{The Seppecher intermediate solution}
\label{sec:seppecher}

In an intermediate region, Seppecher\cite{seppecher} considers two phases of
equal viscosity, and assumes classical incompressible flow, which was
justified here in Sec.~\ref{sec:siout} in the leading order outer equations.
The case when $\Rey=0$ is considered, \ie~the Stokes equations are taken. In
streamfunction formulation, the problem reduces to solving the biharmonic
equation for each phase, thus
\begin{align}
\nabla^4 \Psi_A = 0, \qquad \text{and} \qquad \nabla^4 \Psi_B = 0,\label{eq:BHE}
\end{align}
where phase A is in the region $\Phi<\theta<\pi$, phase B is in the region $0<\theta<\Phi$, and the interface is taken to be located at the apparent angle $\theta=\Phi$. Note that in polar coordinates, the biharmonic operator is defined as
\begin{align}
\nabla^4 =
\frac{\partial^4 }{\partial r^4}
+ \frac{2}{r^2} \frac{\partial^4 }{\partial r^2\partial \theta^2}
+ \frac{1}{r^4} \frac{\partial^4 }{\partial \theta^4}
+ \frac{2}{r} \frac{\partial^3 }{\partial r^3}
- \frac{2}{r^3} \frac{\partial^3 }{\partial r \partial \theta^2}
- \frac{1}{r^2} \frac{\partial^2 }{\partial r^2}
+ \frac{4}{r^4} \frac{\partial^2 }{\partial \theta^2}
+ \frac{1}{r^3} \frac{\partial }{\partial r},
\end{align}
and the velocities in radial and angular directions are
\begin{align}
u_r = \frac{1}{r} \pfrac{\psi}{\theta} \qquad \mbox{and} \qquad u_\theta = - \pfrac{\psi}{r},
\end{align}
respectively. The boundary conditions considered are
\begin{align}
 \label{eq:sep:nfw}
 \left.\Psi_A\right|_{\theta=\pi} &= 0, & \left.\Psi_B\right|_{\theta=0} &= 0,\\
 \label{eq:sep:nsw}
 \left.\pfrac{\Psi_A}{\theta} \right|_{\theta=\pi} &= V_c r, & \left.\pfrac{\Psi_B}{\theta}\right|_{\theta=0} &= -V_c r,\\
 \label{eq:sep:nfi}
 \left.\Psi_A\right|_{\theta=\Phi} &= D_A, & \left.\Psi_B\right|_{\theta=\Phi} &= -D_B,\\
 \label{eq:sep:cty}
 \left.\pfrac{\Psi_A}{\theta}\right|_{\theta=\Phi} &= \left.\pfrac{\Psi_B}{\theta}\right|_{\theta=\Phi}, &
 \left.\ppfrac{\Psi_A}{\theta} \right|_{\theta=\Phi} &= \mu_{B/A}\left.\ppfrac{\Psi_B}{\theta} \right|_{\theta=\Phi},
\end{align}
being no flux and no-slip at the wall in \refe{eq:sep:nfw} and \refe{eq:sep:nsw}, and no flux through the interface and flow into the contact line region being specified by $D_A$ and $D_B$ in regions $A$ and $B$ respectively, in \refe{eq:sep:nfi}. Finally, continuity of tangential velocity and tangential stress at the interface are given in \refe{eq:sep:cty}, where the assumption of equal viscosities made by Seppecher\cite{seppecher} has been generalised to different viscosities, as considered throughout this study, where $\mu_{B/A}=\mu_{B}/\mu_{A}$ gives the ratio of viscosities. Clearly setting $\mu_{B/A}=1$ will then give the equal viscosity solution.

The problem is linear in $V_c$, and in $D_A$ and $D_B$, such that we can solve the equation for $V_c = 0$ and $D_{A,B} = 0$ separately and then superpose both equations.

\subsection{The static solution, $V_c = 0$}

The biharmonic equation \refe{eq:BHE} is solved in the static case $V_{c} = 0$ by
\begin{align}
\Psi_{A}\lr{\theta} = C_{0,A} + C_{1,A} \theta + C_{2,A} \sin(2\theta) + C_{3,A} \cos(2\theta),\label{eq:vc0sol1}\\
\Psi_{B}\lr{\theta} = C_{0,B} + C_{1,B} \theta + C_{2,B} \sin(2\theta) + C_{3,B} \cos(2\theta).\label{eq:vc0sol2}
\end{align}
It is here that an important discrepancy to the work of
Seppecher\cite{seppecher} occurs. The solution form written there is based on
Eq.~(1.3) of the seminal work of Moffatt,\cite{Moffatt} which gives an
incorrect solution of the biharmonic equation $\nabla^4 \psi=0$ for the
specific case $\psi= f_0(\theta)$. The general solution and the other
particular cases in Ref.~\onlinecite{Moffatt} are correct. The incorrect
solution form is used by Seppecher,\cite{seppecher} which to the authors'
knowledge has not been remarked upon previously in the literature. The
coefficients of the correct solutions \refe{eq:vc0sol1}--\refe{eq:vc0sol2}
for general viscosity are given in Appendix \ref{app:coeffssepp}.

\subsection{The `Huh and Scriven' solution $D_{A,B} = 0$}

\begin{figure}[ht]
\centering
	\includegraphics{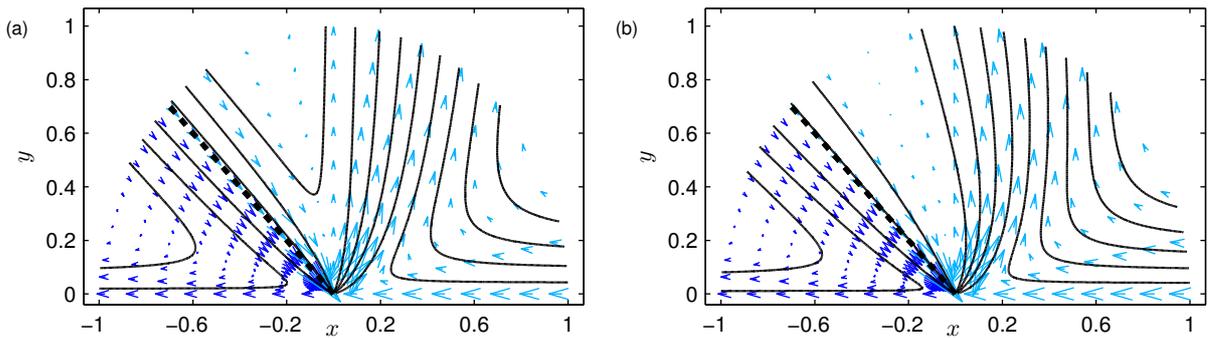}
	\caption{A comparison for a flow scenario in the intermediate region, governed by the Stokes equations, with $\Phi=3\pi/4$, $D_A=0.1$, $D_B=-0.1$, and $V_c=1$ for the two cases (a) $\mu_{B/A}=0$ and (b) $\mu_{B/A}=1$.}
\label{fig:seppflow}
\end{figure}

The biharmonic equation \refe{eq:BHE} in the case $D_{A,B} = 0$ for general viscosities is equivalent to that considered by Huh and Scriven.\cite{HuhScriv71} As such, the solutions are of the form
\begin{align}
\Psi_{A}\lr{r,\theta} = r \lr{ k_{0,A} \sin \theta + k_{1,A}\cos \theta + k_{2,A}\theta \sin\theta + k_{3,A}\theta \cos\theta},\\
\Psi_{B}\lr{r,\theta} = r\lr{ k_{0,B}\sin \theta + k_{1,B}\cos \theta + k_{2,B}\theta \sin\theta + k_{3,B}\theta \cos\theta},
\end{align}
and once again the coefficients are given in Appendix \ref{app:coeffssepp}. A comparison for a flow scenario with $\Phi=3\pi/4$, $D_A=0.1$, $D_B=-0.1$, and $V_c=1$ for the two cases $\mu_{B/A}=0$ and $\mu_{B/A}=1$, corresponding to disparate and equal viscosities respectively, is shown in Fig.~\ref{fig:seppflow}.

%%%%%%%%%%%%%%%%%%%%%%%%%%%%%%%%%%%%%
%%%%%%%%%%%%%%%%%%%%%%%%%%%%%%%%%%%%%
%%%%%%%%%%%%%%%%%%%%%%%%%%%%%%%%%%%%%
%%%%%%%%%%%%%%%%%%%%%%%%%%%%%%%%%%%%%
%%%%%%%%%%%%%%%%%%%%%%%%%%%%%%%%%%%%%
%%%%%%%%%%%%%%%%%%%%%%%%%%%%%%%%%%%%%
%%%%%%%%%%%%%%%%%%%%%%%%%%%%%%%%%%%%%
%%%%%%%%%%%%%%%%%%%%%%%%%%%%%%%%%%%%%
%%%%%%%%%%%%%%%%%%%%%%%%%%%%%%%%%%%%%
%%%%%%%%%%%%%%%%%%%%%%%%%%%%%%%%%%%%%

\section{Contact line region}
\label{sec:cl}

We now consider the inner region near to both interface and wall in polar
coordinates with $r=O(\Cn)$. The scaling $r=\Cn\inin{r}$ (inner variables
denoted with tildes) allows a balance that retains all terms in the governing equations and
boundary conditions, giving a complete dominant balance. As such the contact line scalings are
\begin{gather}
 r = \Cn\inin{r}, \quad \bmN = \Cn^{-1}\inin{\bmN}, \quad
 \rho = \inin\rho, \quad \tbU=\inin\tbU, \quad \tbU_w=\inin\tbU_w, \nonumber\\
 \tens{M} = \Cn^{-1}\inin{\tens{M}}, \quad
 \tens{T} = \Cn^{-1}\inin{\tens{T}}, \quad
 \bmtau = \Cn^{-1}\inin{\bmtau}, \quad
 p = \Cn^{-1}\inin{p},\label{eq:ininscals}
\end{gather}
and variables are then expanded in the sharp-interface limit (for an arbitrary variable $\tilde\phi$) as
\begin{align}
 \tilde{\phi} = \tilde{\phi}^{(0)} + \Cn \tilde{\phi}^{(1)} + \Cn^2 \tilde{\phi}^{(2)} + \ldots,
\end{align}
with the leading order terms being considered, and the superscript $(0)$ dropped for ease of exposition in what is to follow.

\begin{figure}[ht]
\centering
	\includegraphics{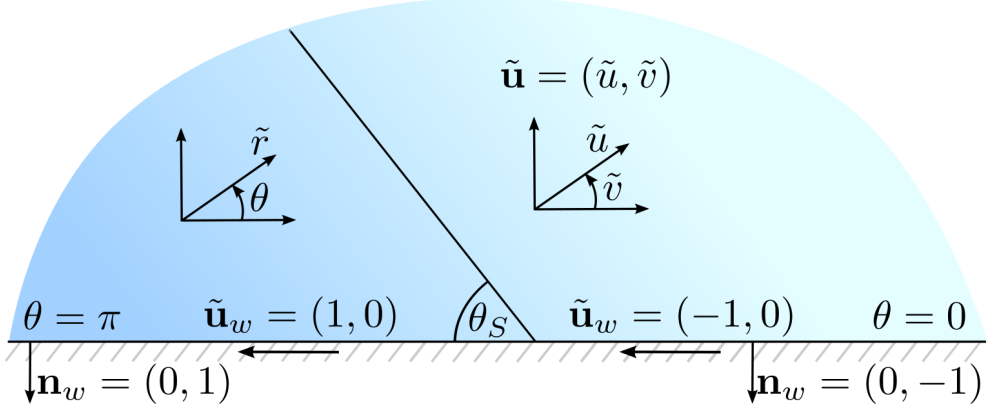}
	\caption{Sketch of the vicinity of the contact line in polar coordinates.}
\label{fig:clregion}
\end{figure}

As the classical formulation of the problem analysed by Huh and Scriven\cite{HuhScriv71} predicted singularities in stress and pressure due to a multivalued velocity at the contact line, the behaviour as the contact line is approached in our diffuse-interface model is of particular interest. In this respect, we consider the (steady) continuity equation
\refe{eq:dfge1} and momentum equations \refe{eq:simpmom}, with boundary conditions \refe{eq:dfgeend}
in polar coordinates, where $\inin u$ and $\inin v$ are the radial and angular velocity components (shown in Fig.~\ref{fig:clregion}), to obtain
\begin{align}
 \pfrac{}{\inin r}(\inin \rho \inin u) + \frac{\inin \rho\inin u}{\inin r} + \frac{1}{\inin r}\pfrac{}{\inin \theta}(\inin \rho \inin v) &= 0, \\
 \frac{1}{\Ca_K}\inin \rho\lr{\pppfrac{\inin \rho}{\inin r} - \frac{1}{\inin r^2}\pfrac{\inin \rho}{\inin r} + \frac{1}{\inin r}\ppfrac{\inin \rho}{\inin r}
 - \frac{2}{\inin r^3}\ppfrac{\inin \rho}{\inin \theta} + \frac{1}{\inin r^2}\pxyyfrac{\inin \rho}{\inin r}{\inin \theta}} -
 \frac{1}{\Ca_K}\pfrac{}{\inin r}\lr{\inin{\rho}^2f'(\inin{\rho})} &\nonumber\\
 + [2\mu(\inin \rho)+\nu(\inin \rho)]\lrsq{\ppfrac{\inin u}{\inin r} + \frac{1}{\inin r}\pfrac{\inin u}{\inin r}
- \frac{\inin u}{\inin r^2}}
+ \frac{\mu(\inin \rho)}{\inin r^2}\ppfrac{\inin u}{\inin \theta}
 - [3\mu(\inin \rho)+\nu(\inin \rho)]\frac{1}{\inin r^2}\pfrac{\inin v}{\inin \theta} + [\mu(\inin \rho)+\nu(\inin \rho)]\frac{1}{\inin r}\pxyfrac{\inin v}{\inin r}{\inin \theta} &\nonumber\\ +
\lrsq{[2\mu'(\inin \rho)+\nu'(\inin \rho)]\pfrac{\inin u}{\inin r}+\nu'(\inin \rho)\frac{1}{\inin r}\lr{\inin u+\pfrac{\inin v}{\inin \theta}}}\pfrac{\inin \rho}{\inin r} + \mu'(\inin \rho)
\lrsq{\frac{1}{\inin r^2}\pfrac{\inin u}{\inin \theta}-\frac{\inin v}{\inin r^2}+\frac{1}{\inin r}\pfrac{\inin v}{\inin r}}\pfrac{\inin \rho}{\inin \theta}
&=0, \\
 \frac{1}{\Ca_K}\inin \rho\frac{1}{\inin r}\lr{\pxxyfrac{\inin \rho}{\inin r}{\inin \theta} + \frac{1}{\inin r}\pxyfrac{\inin \rho}{\inin r}{\inin \theta}
 + \frac{1}{\inin r^2}\pppfrac{\inin \rho}{\inin \theta}} -
\frac{1}{\Ca_K}\frac{1}{\inin r}\pfrac{}{\inin \theta}\lr{\inin{\rho}^2f'(\inin{\rho})} &\nonumber\\
 + \mu(\inin \rho)\lrsq{\ppfrac{\inin v}{\inin r} + \frac{1}{\inin r}\pfrac{\inin v}{\inin r}
- \frac{\inin v}{\inin r^2}}
+ [2\mu(\inin \rho)+\nu(\inin \rho)]\frac{1}{\inin r^2}\ppfrac{\inin v}{\inin \theta}  +
 [3\mu(\inin \rho)+\nu(\inin \rho)]\frac{1}{\inin r^2}\pfrac{\inin u}{\inin \theta} + [\mu(\inin \rho)+\nu(\inin \rho)]\frac{1}{\inin r}\pxyfrac{\inin u}{\inin r}{\inin \theta} &\nonumber\\
+\mu'(\inin \rho)\lrsq{\frac{1}{\inin r}\pfrac{\inin u}{\inin \theta}-\frac{\inin v}{\inin r}+\pfrac{\inin v}{\inin r}}
\pfrac{\inin \rho}{\inin r}+
\lrsq{[2\mu'(\inin \rho)+\nu'(\inin \rho)]\frac{1}{\inin r^2}\lr{\inin u+\pfrac{\inin v}{\inin \theta}}+\nu'(\inin \rho)\frac{1}{\inin r}\pfrac{\inin u}{\inin r}}
\pfrac{\inin \rho}{\inin \theta}
 &=0,
\end{align}
and the boundary conditions
\begin{align}
 \inin{u} &= -1, & \inin{v} &= 0, & -\frac{\Ca}{6\Ca_K}\frac{1}{\inin{r}}\pfrac{\inin{\rho}}{\theta} &= \cos\theta_S(1-\inin{\rho})\inin{\rho}, \qquad &\mbox{on }\theta=0,\label{eq:bccl1}\\
 \inin{u} &= 1, & \inin{v} &= 0, & \frac{\Ca}{6\Ca_K}\frac{1}{\inin{r}}\pfrac{\inin{\rho}}{\theta} &= \cos\theta_S(1-\inin{\rho})\inin{\rho}, \qquad &\mbox{on }\theta=\pi.\label{eq:bccl2}
\end{align}
{Note that to transfer these polar velocity components to Cartesian, $\tilde{u}_{\textrm{Cartesian}} = \tilde{u}\cos\theta - \tilde{v}\sin\theta$, $\tilde{v}_{\textrm{Cartesian}} = \tilde{u}\sin\theta + \tilde{v}\cos\theta$.} To consider the asymptotic solution as the contact line is approached, we consider the expansions
\begin{align}
 \inin\rho &= \inin\rho_0(\theta) + \inin\rho_1(\theta)\inin{r} + \inin\rho_2(\theta)\inin{r}^2 + \cdots, \label{eq:expan1}\\
 \inin{u} &= \inin{u}_0(\theta) + \inin{u}_1(\theta)\inin{r} + \inin{u}_2(\theta)\inin{r}^2 + \cdots, \\
 \inin{v} &= \inin{v}_0(\theta) + \inin{v}_1(\theta)\inin{r} + \inin{v}_2(\theta)\inin{r}^2 + \cdots. \label{eq:expan3}
\end{align}
At leading order we find
\begin{align}
 \inin{u}_0 = -\inin{v}_0' - \frac{\inin{\rho}_0'}{\inin{\rho}_0}\inin{v}_0, \qquad
 \inin{\rho}_0''' = 0, \qquad
 \inin{\rho}_0'' = 0,
\end{align}
subject to $\inin{\rho}_0'=0$ on $\theta=\{0,\pi\}$, $\inin{u}_0=-1$, and
$\inin{v}_0=0$ on $\theta=0$, $\inin{u}_0=1$, and $\inin{v}_0=0$ on
$\theta=\pi$. The general solution for the leading order density is of the
form $\inin{\rho}_0=C_{\inin\rho_1}\inin\theta + C_{\inin\rho_2}$, and
satisfying the wetting boundary condition {at $O(r^{-1})$} forces $C_{\inin\rho_1}=0$. Since
the interface was defined at $\inin{\rho}=\ilfrac{1}{2}$, we thus require
$C_{\inin\rho_2}=\ilfrac{1}{2}$, but leave it general to see its effect, and
so that generalising to other definitions of interface could be considered.
As we cannot solve for the velocities we continue to the following order and
find
\begin{align}
 \inin{u}_0 = -\inin{v}_0', \qquad
 \mu(C_{\inin\rho_2})\frac{\Ca_K}{C_{\inin\rho_2}}(\inin{v}_0''' + \inin{v}_0')
 + \inin{\rho}_1'' + \inin{\rho}_1 = 0 \qquad
 \mu(C_{\inin\rho_2})\frac{\Ca_K}{C_{\inin\rho_2}}(\inin{v}_0'' + \inin{v}_0)
 - \inin{\rho}_1''' - \inin{\rho}_1' = 0.
\end{align}
The general solution to these equations may be found to be
\begin{align}
 \inin{\rho}_1 = A_{\inin\rho_1}\sin\theta + A_{\inin\rho_2}\cos\theta + A_{\inin\rho_3} \theta\sin\theta
 + A_{\inin\rho_4} \theta\cos\theta, \\
 \inin{u}_0 = \frac{C_{\inin\rho_2}}{\mu(C_{\inin\rho_2})\Ca_K}
 \lrsq{(A_{\inin\rho_4}\theta + A_{\inin\rho_5})\cos\theta
 + (A_{\inin\rho_3}\theta + A_{\inin\rho_6})\sin\theta}, \\
 \inin{v}_0 = \frac{C_{\inin\rho_2}}{\mu(C_{\inin\rho_2})\Ca_K}
 \lrsq{(A_{\inin\rho_6}-A_{\inin\rho_4}+A_{\inin\rho_3}\theta)\cos\theta
 - (A_{\inin\rho_5}+A_{\inin\rho_3} + A_{\inin\rho_4}\theta)\sin\theta}.
\end{align}
The no-slip and wetting boundary conditions on $\theta=\{0,\pi\}$ {are from \refe{eq:bccl1}--\refe{eq:bccl2}. Using expansions \refe{eq:expan1}--\refe{eq:expan3}, these} are
\begin{align}
 \inin{u}_0 = \mp1, \qquad \inin{v}_0 = 0, \qquad \mp\frac{1}{6}\frac{\Ca}{\Ca_K}\inin\rho_1' = \cos(\theta_S)C_{\inin\rho_2}(1-C_{\inin\rho_2}),
\end{align}
{at $O(r^0)$}, which then give the density and velocity components as
\begin{align}
 \inin\rho_1 = \frac{6\Ca_K C_{\inin\rho_2} (C_{\inin\rho_2}-1)}{\Ca}\cos\theta_S\sin\theta + A_{\inin\rho_2}\cos\theta, \qquad
 \inin{u}_0 = -\cos\theta,\qquad
 \inin{v}_0 = \sin\theta,
\end{align}
where $A_{\inin\rho_2}$ {is a constant of integration. If we assume that at these very small distances to the contact line the profile is planar (at the Young contact angle $\theta_S$), then we require} $\inin{\rho}(\inin{r},\pi-\theta_S)=C_{\inin\rho_2}$ {(or equivalently $\inin{\rho}_1(\inin{r},\pi-\theta_S)=0$)}, at least up to this first order correction. As such
\begin{align}
 &\frac{6\Ca_K C_{\inin\rho_2} (C_{\inin\rho_2}-1)}{\Ca}\cos\theta_S\sin(\pi-\theta_S) + A_{\inin\rho_2}\cos(\pi-\theta_S)=0\nonumber\\
\Rightarrow\qquad&
A_{\inin\rho_2} = 6\frac{\Ca_K}{\Ca} C_{\inin\rho_2} \sin(\theta_S) (C_{\inin\rho_2}-1),
\end{align}
however, as with $C_{\inin\rho_2}$, we will leave this general to see its effect. We now consider the stresses and pressure, scaled as in \refe{eq:ininscals}. The total stress components in polar coordinates and in inner variables are
\begin{align}
\inin{\tens M}_{{\inin{r}}{\inin{r}}} &=  \frac{1}{\Ca_K}\left[
\frac{1}{2}\lrcur{2\inin\rho f(\inin\rho) + \frac{1}{\inin r^2}\lr{\pfrac{\inin\rho}{\theta}}^2-\lr{\pfrac{\inin\rho}{\inin r}}^2} \right.\nonumber\\&\left. + \inin\rho\lrcur{
\ppfrac{\inin\rho}{\inin r} + \frac{1}{\inin r}\pfrac{\inin\rho}{r} + \frac{1}{\inin r^2}\ppfrac{\inin\rho}{\theta} - \pfrac{}{\inin\rho}(\inin\rho f(\inin\rho))}
\right] + [2\mu(\inin\rho)+\nu(\inin\rho)]\pfrac{\inin u}{\inin r}+\nu(\inin\rho)\frac{1}{\inin r}\lr{\inin u + \pfrac{\inin v}{\theta}},\\
\inin{\tens M}_{\inin{r}\theta} &=  -\frac{1}{\Ca_K}\frac{1}{\inin r}\pfrac{\inin\rho}{\inin r}\pfrac{\inin\rho}{\theta} + \mu(\inin\rho)\lrsq{\frac{1}{\inin r}\lr{\pfrac{\inin u}{\theta} - \inin v} + \pfrac{\inin v}{\inin r}},\\
\inin{\tens M}_{\theta\theta} &=  \frac{1}{\Ca_K}\left[
\frac{1}{2}\lrcur{2\inin\rho f(\inin\rho) - \frac{1}{\inin r^2}\lr{\pfrac{\inin\rho}{\theta}}^2+\lr{\pfrac{\inin\rho}{\inin r}}^2} \right.\nonumber\\&\left. + \inin\rho\lrcur{
\ppfrac{\inin\rho}{\inin r} + \frac{1}{\inin r}\pfrac{\inin\rho}{r} + \frac{1}{\inin r^2}\ppfrac{\inin\rho}{\theta} -
\pfrac{}{\inin\rho}(\inin\rho f(\inin\rho))
}
\right] +  \nu(\inin\rho)\pfrac{\inin u}{\inin r} + [2 \mu(\inin\rho)+ \nu(\inin\rho)]\frac{1}{\inin r}\lr{\inin u + \pfrac{\inin v}{\theta}},
\end{align}
and we find at leading order $\inin{\tens{M}} = O(1)$, as all $O(1/\inin{r})$
terms cancel. To obtain the precise form of the stresses, we require the
second-order terms in the governing equations. The pressure in this inner
region is of the form
\begin{align}
 \inin p = \frac{1}{\Ca_K}\inin{\rho}^2 f'(\inin\rho),
\end{align}
and so at leading order as $\inin r \to 0$, we find that $\inin p = {\Ca_K}^{-1}C_{\inin\rho_2}^2 f'(C_{\inin\rho_2}) + O(\inin r)$, being finite at the contact line.

Continuing to find the second-order terms, we obtain the density and velocity
corrections
\begin{align}
 \inin\rho_2 &=
D_{\inin \rho_1}+D_{\inin \rho_2}\cos(2\theta)
+ \frac{6\Ca_K}{\Ca} A_{\inin\rho_2} \cos\theta_S (C_{\inin\rho_2}-1/2)\sin(2\theta),
\\
\lr{\begin{array}{c} \inin{u}_1 \\ \inin{v}_1 \end{array}} &=
\lr{\begin{array}{c}
\frac{A_{\inin\rho_2}}{2C_{\inin\rho_2} }(1-\cos(2\theta))
- D_{\inin v}\sin(2\theta)\\
\frac{A_{\inin\rho_2}}{2C_{\inin\rho_2}}\sin(2\theta)+ D_{\inin v}(1-\cos(2\theta))
\end{array}}
,
\end{align}
where the arbitrary constants $D_{\inin \rho_1}$, $D_{\inin \rho_2}$ and
$D_{\inin v}$ would be set by {a full numerical solution of the contact line region. Using the results of the
velocity, and converting into Cartesian components, we see that at the contact line
\begin{align}
\tilde{u}_{\textrm{Cartesian}} = -1 -2 \tilde{y} D_{\inin v} + \cdots, \qquad \tilde{v}_{\textrm{Cartesian}} = \frac{D_{\inin v}}{C_{\inin\rho_2}} \tilde{y} + \cdots,
\end{align}
showing the well-defined velocity approaching the contact line, at the wall $\tilde{y}=0$.}
All of these results allow us to determine the leading-order stress components as
\begin{align}
 \inin{\tens M}_{\inin{r}\inin{r}} &=
 M_1\cos(2\theta) -
 M_2\sin(2\theta) + M_3 + O(\inin r),
 \\
 \inin{\tens M}_{\inin{r}\inin{\theta}} &=
 -M_2\cos(2\theta) - M_1\sin(2\theta) + O(\inin r),
 \\
 \inin{\tens M}_{\inin{\theta}\inin{\theta}} &= M_2\sin(2\theta)
 -M_1\cos(2\theta) + M_3 + O(\inin r),
\end{align}
where
{\begin{align}
M_1 &= \frac{9\Ca_K(1-C_{\inin\rho_2})^2C_{\inin\rho_2}^2}{\Ca^2}(1+\cos(2\theta_S))
-\frac{A_{\inin\rho_2}}{C_{\inin\rho_2}}\mu(C_{\inin\rho_2})
-\frac{1}{2\Ca_K}A_{\inin\rho_2}^2
\\
M_2 &= -6\frac{A_{\inin\rho_2}C_{\inin\rho_2}}{\Ca}\cos(\theta_S)
(1-C_{\inin\rho_2})
+ 2\mu(C_{\inin\rho_2})C_{\inin v},\\
M_3 &= \frac{A_{\inin\rho_2}}{C_{\inin\rho_2}}[\mu(C_{\inin\rho_2})+\nu(C_{\inin\rho_2})]
- \frac{1}{\Ca_K}\lrsq{C_{\inin\rho_2}^2f'(C_{\inin\rho_2}) -4D_{\inin\rho_1}C_{\inin\rho_2}},
\end{align}}%
showing that the stresses are nonsingular as $\inin r \to 0$, as all $O(\inin r^{-1})$ terms cancel.

{Recall from Sec.~\ref{sec:silim} where we
considered the sharp-interface limit away from the walls, that the
(liquid-gas) interface gives rise to the classical boundary conditions
(kinematic boundary condition, continuity of velocity, normal and tangential
stress balances) in this limit, and that the two bulk phases satisfy the
Navier--Stokes equations. This cannot be extended to the contact line region
as the very reason that the diffuse-interface model resolves the moving
contact line singularities is the diffuse nature (and finite thickness) of
the interface. As already noted at the start of Sec.~\ref{sec:cl}, the
dominant balance of scalings for the contact line region shows that this
region is of size $O(\Cn)$, so the region where a diffuse-interface (i.e. not
a sharp one) is significant is of this order (and as mentioned, all terms in
the governing equations are retained). In fact the limit $\Cn \to 0$ is a
singular perturbation problem, as setting $\Cn = 0$ gives back the unsolvable
classic problem (without any additional ingredients such as slip).}

\section{Conclusions}
\label{sec:conclusions}

We have undertaken an analytical investigation of the contact line behaviour
for a liquid-gas system with two basic elements: (a) the interface has a
finite thickness, as expected from statistical mechanics studies, and (b) the
no-slip condition is applied at the wall. The variation of the two
viscosities between the liquid and gas phase have been left as general
functions so as to be relevant to both equal (constant) viscosities, as
considered by Seppecher,\cite{seppecher} and for disparate viscosities. The
form of the double well contribution to the free energy has also been left
general. In this case, our diffuse-interface model is then shown to resolve
the singularities associated with the moving contact line problem without the
need for any further physical effects from the microscale.

To summarize, our main results are:
\begin{itemize}
 \item Determined the governing equations, identifying a suitable wall free energy form, in \refe{eq:wallfdim}, to give a boundary condition to isolate the effect of the diffuse-interface on the contact line behaviour, without density gradients at the wall at large distances from the contact line.
 \item Considered a general viscosity relationship between the two fluid phases.
 \item Justified the model equations through considering the sharp-interface limit away from the contact line region, obtaining the classical bulk equations and interface conditions.
 \item Highlighted and corrected a discrepancy in the `intermediate region' of the seminal work of Seppecher.\cite{seppecher}
 \item Shown that the diffuse-interface model we adopted resolves the
     singularities associated with the moving contact line problem by
     considering the asymptotic behaviour in the vicinity of the contact
     line.
\end{itemize}

The model studied here imposed no-slip and enforced a wetting boundary
condition causing the wall free energy to relax to equilibrium
instantaneously. These conditions may be altered to allow for additional
effects, if they are considered physically relevant, although doing so is not
necessary to resolve the moving contact line problem, as determined in this
study. Appropriate generalisations of the boundary conditions at the wall are
to allow for finite-time relaxation, and to apply the generalised Navier
boundary condition (GNBC) to allow for slip at the wall, these being
discussed and derived by Qian, Wang, and
Sheng,\cite{QianWangShengGNBCfirst,QianWangShengJFM} Jacqmin,\cite{jacqmin}
and Yue and Feng,\cite{YueFengEPJ,YueFengPoF} {for binary fluids}. In dimensional form and for our
liquid-gas configuration these are {instead given by}
\begin{equation}
 \lambda\lr{\pilfrac{\bar\rho}{\bar t} + \bar\tbU\bmDot\bar\bmN\bar\rho} = -\bar L(\bar\rho),\label{eq:wallrelaxwet}
\end{equation}
where $ \bar L(\bar\rho) = K \tbN_w \bmDot \bar\bmN \bar\rho + \bar f'_w(\bar\rho)$ is the wall chemical potential, and with $\lambda=0$ representing instantaneous relaxation to equilibrium, and
\begin{equation}
 \bar L(\bar\rho) (\vect{t}_w\bmDot\bar\bmN\bar\rho) - \bar\bmtau_{nt} = \bar\beta(\bar\tbU-\bar\tbU_w)\bmDot\vect{t}_w,
\end{equation}
where $\bar\beta$ is the inverse slip length, $\vect{t}_w$ the tangent to the
wall, and $\bar\bmtau_{nt}$ is the viscous shear stress. Note that
$\lambda=0$ reduces to the popular Navier-slip condition. It is also of
interest to note that such boundary conditions allow the diffuse-interface
model to have a velocity dependent microscopic contact angle and to exhibit
effective slip on the solid, both features of the continuum mechanical
interface formation model of Shikhmurzaev\cite{ShikhBook} (and analysed in
Ref.~\onlinecite{My_Shikh}).

The analysis presented here does not include boundary layers at the wall,
which would arise if the wetting boundary condition is replaced by conditions
such as \refe{eq:wallrelaxwet}, or a form to model a precursor film such as
that used by Pismen and Pomeau\cite{PismenPomeau}, where $\bar{\rho}=\rho_a$,
a constant, on $\partial\Omega$. This would be a rather involved problem to
investigate using matched asymptotics, as the boundary layer at the wall
would meet the boundary layer at the liquid-gas interface, and the two would
need to match when ``turning the corner'' at the contact line, where the full
inner equations should be retained, and most likely matching would have to be
done numerically.

We believe that the present study will motivate further analytical and
numerical work with diffuse-interface models,
{not just for ideally smooth substrates as
was the case here but for (chemically or topographically) heterogeneous ones where
previous studies have utilized a sharp-interface model with a slip boundary condition,\cite{Savva09,Savva10,SavvaPavliotisKalliadasis1,SavvaPavliotisKalliadasis2,RajChemHet,Herde12}
or have been numerically investigated with phase-field models.\cite{DupuisYeomans,KusumaatmajaYeomans,christophe_marc}
}
The binary fluid moving contact line problem
{(\ie~for a solid-liquid-liquid problem)}
is certainly of interest, but there remains ongoing debate on
the correct scaling in the sharp-interface limit for an additional
nondimensional ``mobility'' parameter, associated with the additional
Cahn--Hilliard type equation for the order
parameter.\cite{KhatavkarChemEngSci,YueZhouFeng,AbelsGarckeGrun12,Magaletti_etal_2013}
There
has, however, been some early progress in the work of
Ref.~\onlinecite{WangWangSharp}.

{Additionally, the question of how diffuse-interface models may be adapted for non-Newtonian fluids is of interest. It is known that certain forms of shear-thinning non-Newtonian models are able to resolve the moving contact line problem without features like a diffuse-interface and whilst still applying no-slip,\cite{WeidnerSchwartz} however there are obviously a wealth of non-Newtonian behaviours other than shear-thinning which may be considered (for instance viscoelastic fluids, modelled with Maxwell or Oldroyd models). For an \emph{incompressible binary fluid model}, a numerical investigation of an Oldroyd-B fluid in a Newtonian solvent has been studied, \cite{YueetalGooch} with the stress tensor of the Oldroyd-B fluid coupled into the total stress tensor. A possible extension to our analysis would be the consideration of a similar Oldroyd-B model, which would be in compressible form and coupled into our system of equations by replacing \refe{eq:origM} in the manuscript by
\begin{gather}
 \bar{\textbf{M}} = \bar{\textbf{T}} + \bar{\bm{\tau}} + \bar{\bm{\tau}}_{\textrm{poly}}, \qquad
 \bar{\bm{\tau}}_{\textrm{e}} = \bar{\bm{\tau}}_{\textrm{poly}}
 + \bar{\nu}(\rho)[(\bar{\bm{\nabla}}\bm{\cdot}\bar{\textbf{u}})\textbf{I}], \nonumber\\
 \bar{\bm{\tau}}_{\textrm{e}} + \bar\lambda_{\textrm{e}}\stackrel{\triangledown}{\bar{\bm{\tau}}}_{\textrm{e}}  \ =
 \left[ \bar\mu_{\textrm{e}} + \frac{\upd \bar\mu(\bar{\rho})}{\upd \bar\rho} \bar\lambda_{\textrm{e}} \, \bar\rho \,
 (\bar{\bm{\nabla}}\bm{\cdot}\bar{\textbf{u}}) \right]
 \left[(\bar{\bm{\nabla}}\bar{\textbf{u}})+(\bar{\bm{\nabla}}\bar{\textbf{u}})^\mathrm{T}\right],
\end{gather}
where $\bar{\bm{\tau}}_{\textrm{poly}}$ is the stress contribution from the polymer (so the usual Oldroyd stress tensor including polymer and solvent stresses is $\bm\tau_o = \bar{\bm{\tau}} + \bar{\bm{\tau}}_{\textrm{poly}}$),
$\bar{\bm{\tau}}_{\textrm{e}}$ is a stress tensor chosen for convenience in writing the governing equations---effectively being $\bm\tau_o$ minus an \emph{incompressible} Newtonian stress tensor,
$\bar\mu_{\textrm{e}}$ is the polymer viscosity, and $\bar\lambda_{\textrm{e}}$ is the stress relaxation time (both of which could also be allowed to be density dependent in the above representation). These equations may be derived from
equations \refe{eq:origcty}, \refe{eq:origtau} and the usual Oldroyd-B model
\begin{align}
\bm\tau_o + \bar\lambda_{\textrm{e}}\stackrel{\triangledown}{\bm\tau}_o = 2(\bar\mu_{\textrm{e}} + \bar\mu)(\textbf{D} + \zeta_o\stackrel{\triangledown}{\textbf{D}}), \qquad \textbf{D} = \left[(\bar{\bm{\nabla}}\bar{\textbf{u}})+(\bar{\bm{\nabla}}\bar{\textbf{u}})^\mathrm{T}\right]/2,
\end{align}
where $\zeta_o = \bar\mu\bar\lambda_{\textrm{e}}/(\bar\mu_{\textrm{e}} + \bar\mu)$ is the retardation time of the Oldroyd-B model. Note here that the triangle denotes the upper convected stress derivative, defined for a general tensor $\bar{\textbf{A}}$ as
\begin{align}
 \stackrel{\triangledown}{\bar{\textbf{A}}} = \frac{\partial \bar{\textbf{A}}}{\partial \bar{t}}
 + \bar{\textbf{u}}\bm{\cdot}\bar{\bm{\nabla}}\bar{\textbf{A}} - (\bar{\bm{\nabla}}\bar{\textbf{u}})\bar{\textbf{A}}
 - \bar{\textbf{A}}(\bar{\bm{\nabla}}\bar{\textbf{u}})^{\textrm{T}}.
\end{align}
}

Of particular interest for future work would also be the inclusion of
non-local terms into the governing equations. This is considered for
equilibrium wetting using a {relatively simple
density-functional formulation} by Pereira and Kalliadasis,\cite{antoniojfm}
capturing the non-local effects due to long-range intermolecular
interactions.
{More involved density-functional theories, including
inertial effects or hydrodynamic interactions, or both, are currently being
pursued.\cite{RexLowenEPJE,BenPRL,BenJPCM} The analysis here cannot be
easily extended to include the relevant non-local terms, as dealing with
integrals over the entire fluid volume in an asymptotic procedure (such as
taking the sharp-interface limit) presents a highly nontrivial task.}

%%%%%%%%%%%%%%%%%%
%
% ACKNOWLEDGEMENTS
%
%%%%%%%%%%%%%%%%%%

\section*{Acknowledgements}
We are grateful to the anonymous Referees and
Dr.~Marc Pradas for useful comments and suggestions. We acknowledge
financial support from ERC Advanced Grant No. 247031 and Imperial College
through a DTG International Studentship.

%%%%%%%%%%%%
%
% APPENDICES
%
%%%%%%%%%%%%

\appendix

{
\section{Square gradient approach from density functional theory}
\label{app:sqdf}}
{The diffuse-interface free energy in Eq.~\refe{eq:origF} may be derived through a Taylor series expansion of the density from a non-local model,
which is also used in equilibrium density functional theory calculations, as
suggested in Sec.~\ref{sec:probspec}. We give details here for the interested
reader, noting that similar derivations are shown in
Refs.~\onlinecite{CH58,Pismenmeso,antoniojfm}.}

{In equilibrium density functional theory, the grand potential $\Omega^{\textrm{DFT}}$ of the system is given
(\eg~see Refs.~\onlinecite{EvansReview,antoniojfm,PetrPRE}) by
\begin{align}
 \Omega^{\textrm{DFT}}[\bar\rho] =
 \int \fhs(\bar\rho) \nr \dr + \frac{1}{2} \iint \nr\nrp \pattr(\rmrp) \drp\dr + \int \nr(\Vr - \bar{G}) \dr,
\end{align}
where $\nr$ is the position dependent density, $\fhs(\bar\rho)$ is the local
hard sphere free energy, $\Vr$ is the non-local wall potential, and
$\pattr(\rmrp)$ is the interaction potential between two particles. This
grand potential uses a local density approximation (LDA) for hard-sphere
interactions, with a mean-field theory for the attractive perturbation
modelled using the Barker and Henderson approach:\cite{barker_henderson}
\begin{align}
\pattr(\bar{r}) = \left\{ \begin{array}{ll} 0 & \bar{r}\leq\hat\sigma \\
 4\hat\epsilon\lrsq{ \lr{\frac{\hat\sigma}{\bar{r}}}^{12} - \lr{\frac{\hat\sigma}{\bar{r}}}^{6} } & \bar{r}>\hat\sigma
 \end{array} \right. ,
\end{align}}%
{where $\hat\sigma$ is the soft-core parameter of the Lennard-Jones potential of the fluid-fluid interaction,
and $\hat\epsilon$ is a parameter that measures the strength of the
potential. At equilibrium, the grand potential is minimised:
\begin{equation}
 \frac{\delta \Omega^{\textrm{DFT}}[\bar\rho]}{\delta \nr} = 0,  \qquad \mbox{giving} \qquad
 \pfrac{}{\bar\rho}(\bar\rho\fhs(\bar\rho)) + \int \nrp \pattr(\rmrp)\drp + \Vr = \bar{G},
\end{equation}}%
{the non-locality involving the interaction potential being clear. Obtaining the square gradient approximation,
a Taylor series expansion of $\nr$ gives
\begin{align}
\bar{\mathscr{F}}[\bar\rho] = \int (\fhs(\bar\rho)-\bar{G}-\alpha\nr) \nr \dr + \int \frac{K}{2}\lrmod{\bar\bmN\nr}^2\dr + \int \bar{f}_w(\bar\rho) \upd A,\label{eq:FEfromDFT}
\end{align}
neglecting terms of the third derivative and higher, and thus assuming the
density of the fluid varies slowly, as for instance a fluid near its critical
point. The non-local wall potential $\Vr$ has been replaced by a wall free
energy $\bar{f}_w(\bar\rho)$ on the substrate, and coefficients $K$ and
$\alpha$ are given by
\begin{align}
  K = -\frac{2\pi}{3}\int \bar{r}^4\pattr(\bar{r})\upd \bar{r} = \frac{9}{7}\alpha\hat{\sigma}^2 , \qquad \mbox{and} \qquad \alpha = -\frac{1}{2}\iiint\pattr(\lrmod{\tbR}) \dr = \frac{16}{9}\pi\hat{\epsilon}\hat{\sigma}^3.
\end{align}
By writing $\bar{f}(\bar\rho) = \fhs(\bar\rho) - \alpha\nr$, the free energy
in \refe{eq:FEfromDFT} then reduces to that of the diffuse-interface model in
Eq.~\refe{eq:origF}.\\}

\section{Body fitted coordinates}
\label{app:bfcoordinates}

In Sec.~\ref{sec:siin} we utilize body fitted coordinates $(\bfone,\bftwo)$ (as, for example in Refs.~\onlinecite{AndersonSI,Pego89,CaginalpFife,Lowengrub98,Folchsharp}) to the interface (given by $\vect{r}=\vect{r}_0(\bfone,t)$, where $\vect{r}$ is any arbitrary point). $\bfone$ measures distance along the interface (like an arc length), and $\bftwo$ measures the distance in the normal direction from the point $(x,y)\in\Omega$ to the interface $\Gamma=\Gamma(t)$. We note that these coordinates will only be valid locally in $\bftwo$, i.e. near the interface, as $\bftwo$ can become singular for a curved interface, as it reaches the radius of curvature. For slowly moving contact line problems the distortion of the interface will be small and no issues will arise. From our definition of the coordinates, we have
\begin{align}
 \vect{r} = \vect{r}_0(\bfone,t) + \bftwo \vect{n},
\end{align}
where $\tbN$ is the unit normal to the interface (noting that $\tbN=\bmN \bftwo$). As $\vect{r}$ is an arbitrary point, then the conversion between Cartesian and body fitted coordinates follows
\begin{align}
 \vect{r} = x\vect{i} + y\vect{j} = \bfone\vect{e}_1 + \bftwo \vect{e}_2, \nonumber\\
 \vect{e}_1 = \pfrac{x}{\bfone}\vect{i} + \pfrac{y}{\bfone}\vect{j}, \nonumber\\
 \vect{e}_2 = \pfrac{x}{\bftwo}\vect{i} + \pfrac{y}{\bftwo}\vect{j}.\label{eq:body-cart-basis}
\end{align}
We define $\alpha$ as the angle from $\vect{i}$ to $\vect{e}_1$, so that the curvature $\kappa=\pilfrac{\alpha}{\bfone}$. At $\vect{r}_0$ on the interface we have $X$ and $Y$ being the specific values of $x$ and $y$ (i.e. at a specific $\bfone$). Moving a small amount $d\bfone$ along the interface gives $d\bfone\cos\alpha = dX$ and $d\bfone\sin\alpha=dY$. Now consider $(x,y)$ being a point not on the interface, and at a signed distance $\bftwo$ away. Then we have $x = X-\bftwo\sin\alpha$ and $y=Y+\bftwo\cos\alpha$. Returning to \refe{eq:body-cart-basis} we find
\begin{align}
 \vect{e}_1 &= \left(\pfrac{X}{\bfone}-\bftwo\cos\alpha\pfrac{\alpha}{\bfone}\right)\vect{i} +
 \left(\pfrac{Y}{\bfone}-\bftwo\sin\alpha\pfrac{\alpha}{\bfone}\right)\vect{j} =
 (1-\kappa\bftwo)(\cos\alpha\vect{i}+\sin\alpha\vect{j}) \nonumber\\ &= (1-\kappa\bftwo)\vect{t}, \\
 \vect{e}_2 &= \left(\pfrac{X}{\bftwo}-\sin\alpha\right)\vect{i} +
 \left(\pfrac{Y}{\bftwo}+\cos\alpha\right)\vect{j} = -\sin\alpha\vect{i}+\cos\alpha\vect{j} \nonumber\\ &= \tbN,
\end{align}
giving the basis vectors for the body fitted coordinates, where $\vect{t}$ is the unit tangent to the interface.

The components of the metric tensor of these curvilinear coordinates are used to determine operators such as the gradient and Laplacian in these coordinates, and are found through $g_{ij}=\vect{e}_i\bmDot\vect{e}_j$. For orthogonal curvilinear coordinates (such as we have here) we also have $\vect{e}^1 = g^{11}\vect{e}_1 + g^{12}\vect{e}_2$, $\vect{e}^2 = g^{12}\vect{e}_1 + g^{22}\vect{e}_2$ and $g^{11}=1/g_{11}$, $g^{22}=1/g_{22}$, $g^{12}=g_{12}=0$, thus
\begin{align}
 g_{11} = (1-\kappa\bftwo)^2, \quad g_{12} = 0, \quad g_{22} = 1, \mbox{ and }
 g^{11} = (1-\kappa\bftwo)^{-2}, \quad g^{12} = 0, \quad g^{22} = 1.
\end{align}
Having this information, we are now able to determine the relevant terms in the governing equations. Time derivatives satisfy
\begin{align}
 \pfrac{}{t}\rho(x,y,t) = \pfrac{}{t}{\rho}(\bfone,\bftwo,t) + \tbU^s\bmDot\bmN{\rho}(\bfone,\bftwo,t)
\end{align}
where $\tbU^s=(\pilfrac{\bfone}{t},\pilfrac{\bftwo}{t})$ is the interface velocity, showing that the time derivative in the comoving frame is a material derivative. Considering $\tbU=({u},v)$, and writing $h = \sqrt{g_{11}g_{22}}$ for convenience, then
\begin{align}
 \bmN \rho &= \pfrac{\rho}{\bfone}\vect{e}^1 + \pfrac{\rho}{\bftwo}\vect{e}^2 = (1-\kappa\bftwo)^{-2}\pfrac{\rho}{\bfone}\vect{e}_1
 + \pfrac{\rho}{\bftwo}\vect{e}_2, \\
 |\bmN \rho|^2 &= (1-\kappa\bftwo)^{-2}\left(\pfrac{\rho}{\bfone}\right)^2 + \left(\pfrac{\rho}{\bftwo}\right)^2, \\
 \bmN \bmDot \tbU &= \frac{1}{h}\left[ \pfrac{}{\bfone}\left( h {u} \right)
 + \pfrac{}{\bftwo}\left( h v \right) \right] = \pfrac{v}{\bftwo} - \frac{\kappa v}{1-\kappa\bftwo} +
 \pfrac{{u}}{\bfone} - \frac{{u}\bftwo}{1-\kappa\bftwo}\pfrac{\kappa}{\bfone},\\
 \nabla^2\rho &= \frac{1}{h}\left[ \pfrac{}{\bfone}\left( \frac{h}{g_{11}} \pfrac{\rho}{\bfone} \right)
 + \pfrac{}{\bftwo}\left( \frac{h}{g_{22}} \pfrac{\rho}{\bftwo} \right) \right]  \nonumber\\ &=
 \frac{1}{(1-\kappa\bftwo)^2}\ppfrac{\rho}{\bfone} + \frac{\bftwo\pilfrac{\kappa}{\bfone}}{(1-\kappa\bftwo)^3}\pfrac{\rho}{\bfone}
 + \ppfrac{\rho}{\bftwo} - \frac{\kappa}{1-\kappa\bftwo}\pfrac{\rho}{\bftwo}.
\end{align}
For certain matrix quantities the Christoffel symbols of the second kind will be required, which are defined as
\begin{align}
 \Gamma_{ij}^k = \frac{g^{km}}{2}\left( \pfrac{g_{mi}}{\chi_j}+\pfrac{g_{mj}}{\chi_i} + \pfrac{g_{ij}}{\chi_m}\right),
\end{align}
where $\chi_1 = \bfone$ and $\chi_2 = \bftwo$, so that
\begin{align}
 \Gamma_{11}^1 = -\frac{\bftwo}{1-\kappa\bftwo}\pfrac{\kappa}{\bfone}, \quad \Gamma_{12}^1 =
 \Gamma_{21}^1 = -\frac{\kappa}{1-\kappa\bftwo}, \quad \Gamma_{11}^2 = \frac{\kappa}{1-\kappa\bftwo}, \quad \mbox{and}
 \quad \Gamma_{ij}^k = 0 \mbox{ otherwise}.\label{eq:christoffel}
\end{align}
We may then determine
\begin{align}
 \tens{I} &= \vect{e}^1\otimes\vect{e}_2 + \vect{e}^2\otimes\vect{e}_1 = \left( \begin{array}{cc} (1-\kappa\bftwo)^{-2} & 0
 \\ 0 & 1 \end{array} \right),\\
 \nabla\rho\otimes\nabla\rho &= \left( \begin{array}{cc} \left[(1-\kappa\bftwo)^{-2}\pilfrac{\rho}{\bfone}\right]^2 &
 (1-\kappa\bftwo)^{-2}(\pilfrac{\rho}{\bfone})(\pilfrac{\rho}{\bftwo}) \\ (1-\kappa\bftwo)^{-2}(\pilfrac{\rho}{\bfone})(\pilfrac{\rho}{\bftwo})
 & \left( \pilfrac{\rho}{\bftwo} \right)^2 \end{array} \right),\\
 \bmN\tbU &= \left( \pfrac{u_i}{\chi_k} + \Gamma^i_{lk}u_l \right)\vect{e}_i\otimes\vect{e}^k \nonumber\\ &=
 \left( \begin{array}{cc} \pilfrac{{u}}{\bfone} - \bftwo(\pilfrac{\kappa}{\bfone})(1-\kappa\bftwo)^{-1}{u}
 -\kappa(1-\kappa\bftwo)^{-1}v & \pilfrac{u}{\bftwo} - \kappa(1-\kappa\bftwo){u} \\ \pilfrac{v}{\bfone} +
  \kappa(1-\kappa\bftwo){u} & \pilfrac{v}{\bftwo} \end{array} \right),\label{eq:nabu}
\end{align}
where $(u_1,u_2)=(u,v)$, and leading to the components of the viscous stress tensor being
\begin{align}
 (\bmtau)_{11} &=
 \left(\pfrac{{u}}{\bfone}-\frac{\kappa v}{1-\kappa\bftwo} - \frac{\bftwo {u}}{1-\kappa\bftwo}\pfrac{\kappa}{\bfone}\right)
 \left( 2\mu(\rho) + \nu(\rho)(1-\kappa\bftwo)^{-2} \right) + \nu(\rho)(1-\kappa\bftwo)^{-2}\pfrac{v}{\bftwo} \nonumber\\
 (\bmtau)_{12} &= \mu(\rho)\lrsq{\pfrac{{u}}{\bftwo} + \pfrac{v}{\bfone} } \nonumber\\
 (\bmtau)_{22} &= \lr{2\mu(\rho) + \nu(\rho)}\pfrac{v}{\bftwo} + \nu(\rho)\left(\pfrac{{u}}{\bfone} - \frac{\kappa v}{1-\kappa\bftwo} -
 \frac{\bftwo {u}}{1-\kappa\bftwo}\pfrac{\kappa}{\bfone}\right),
\end{align}
and the divergence of our (symmetric) total stress tensor $\tens{M}$ is
\begin{align}
 \bmN\bmDot\tens{M} = \left( \begin{array}{c} \pilfrac{[(\tens{M})_{11}]}{\bfone}+\pilfrac{[(\tens{M})_{12}]}{\bftwo} + 2\Gamma_{11}^1(\tens{M})_{11} +
 3\Gamma_{12}^1(\tens{M})_{12} \\ \pilfrac{[(\tens{M})_{12}]}{\bfone}+\pilfrac{[(\tens{M})_{22}]}{\bftwo} + \Gamma_{11}^1
 (\tens{M})_{12} + \Gamma_{12}^1 (\tens{M})_{22} + \Gamma_{11}^2 (\tens{M})_{11} \end{array} \right).
\end{align}

\section{Coefficients for the Seppecher intermediate solution}
\label{app:coeffssepp}

The intermediate solution of Seppecher was discussed in Sec.~\ref{sec:seppecher}. The coefficients of the solutions of that section are for $V_c=0$:
\begin{align}
 C_{0,A} &= \frac {\cos \Phi}{D_{D0}} \lrsq{ (D_A-D_B) \sin^2\Phi
- 2 \pi \Phi D_A \cos ( 2\Phi)}  ( 1- \mu_{B/A})
\nonumber\\&+ \frac{\sin \Phi} {D_{D0}} \lrsq{ 2 (\pi -\Phi) D_A \cos^2 \Phi + ( D_A+ D_B) \Phi-\pi
 D_B\cos ( 2\Phi)  } ( 1- \mu_{B/A}) \nonumber\\&+
\frac {\sin \Phi}{D_{D0}} \lrsq{  ( D_A+D_B) ( \cos\Phi \sin\Phi -2\pi \sin^2 \Phi -\Phi ) +\pi D_B
  } ,
\\
C_{1,A}&= \frac{2N_{D0}}{D_{D0}},\\
C_{2,A}&= -\frac{N_{D0}}{D_{D0}},\\
C_{3,A}&= \frac{1}{D_{D0} \sin\Phi }
\lrsq{ (\Phi\cos(2\Phi) -\cos\Phi\sin\Phi) D_A - (\Phi-\pi)\sin^2\Phi D_B }  ( 1-\mu_{B/A} ) \nonumber\\&-
\frac {1}{D_{D0} \sin\Phi }
\lrsq{ N_{D0} \cos\Phi - \sin^2 \Phi \lr{  ( D_A+D_B) \Phi-\pi D_B } }
,\nonumber\\
\end{align}
and
\begin{align}
 C_{0,B} &=
-\frac {  [  (\pi -\Phi) \cos \Phi +\sin \Phi ]D_B\sin (2\Phi) ( 1- \mu_{B/A}) }{ D_{D0} } \nonumber\\&
-\frac{
D_A\Phi-
( D_A+D_B) \sin \Phi\cos \Phi + ( \Phi-\pi )  D_B  } { D_{D0} } \sin \Phi ,\\
 C_{1,B} &=
2{\frac { \left[  ( \pi-\Phi) \cos \Phi +\sin \Phi  \right] \cos ( 2\Phi) D_B ( 1-\mu_{B/A}) + \sin^3\Phi(D_A+D_B) }{D_{D0}}},\\
C_{2,B} &= -\frac{1}{2}C_{1,B},\\
C_{3,B} &= -C_{0,B},
\end{align}
where
\begin{align}
 N_{D0} &=  \left\{ D_A \left[ \Phi\cos \Phi \cos ( 2\Phi) -\sin \Phi  \right] + \sin^3 \Phi ( D_A-D_B )
 \right\} ( 1-\mu_{B/A} ) + \sin^3\Phi ( D_A+D_B), \\
 D_{D0} &=
\cos\Phi\left[2 \Phi \lrcur{( \Phi-\pi ) \cos( 2\Phi) +\tan \Phi} + ( \pi-3\Phi ) \sin(2\Phi) +2\sin^2\Phi
\right] ( 1-\mu_{B/A}) -2\pi \sin^3 \Phi.
\end{align}
For $D_{A,B}=0$ the coefficients are:
\begin{align}
 k_{0,A} &= -V_c - \pi k_{2,A} +  k_{1,A}/\pi,\\
 k_{1,A} &= V_c \pi \sin\Phi\cos\Phi \frac{
\mu_{B/A}(\sin^2\Phi-(\Phi-\pi)\Phi-\pi\tan\Phi)+\Phi^2-\sin^2\Phi
}{D_{V0}}
,\\
 k_{2,A} &= V_c \sin^2\Phi \frac{
-\mu_{B/A}(\sin^2\Phi-\Phi^2+\pi\Phi)-\Phi^2+\sin^2\Phi
}{D_{V0}}
,\\
 k_{3,A} &= -\frac{k_{1,A}}{\pi},
\end{align}
and
\begin{align}
 k_{0,B} &= -V_c - k_{3,B} ,\\
 k_{1,B} &= 0
,\\
 k_{2,B} &= -V_c \sin^2\Phi \frac{
\mu_{B/A}(\sin^2\Phi-(\Phi-\pi)^2)+(\Phi-\pi)\Phi-\sin^2\Phi
}{D_{V0}}
,\\
 k_{3,B} &= -V_c \sin\Phi\cos\Phi \frac{
\mu_{B/A}(\sin^2\Phi-(\Phi-\pi)^2)+(\Phi-\pi)\Phi-\sin^2\Phi-\pi\tan\Phi
}{D_{V0}},
\end{align}
where
\begin{align}
 D_{V0} &= \mu_{B/A}(\sin^2\Phi-(\pi-\Phi)^2)(\cos\Phi\sin\Phi-\Phi)+(\sin^2\Phi-\Phi^2)(\Phi-\pi-\cos\Phi\sin\Phi).
\end{align}
For the specific case considered by Seppecher,\cite{seppecher} where $\mu_{B/A}=1$, these substantially simplify. For $V_c=0$, the solutions are
\begin{align}
\Psi_A\lr{\theta} &= \frac{D_A + D_B}{\pi} \lr{ \pi - \theta + \frac{\sin \theta}{\sin \Phi} \sin \lr{\Phi - \theta} + \Phi \frac{\sin^2 \theta}{\sin^2 \Phi}  }
- D_B \frac{\sin^2 \theta }{\sin^2 \Phi}, \\
\Psi_B\lr{\theta} &= \Psi_A\lr{\theta} - \lr{ D_A + D_B},
\end{align}
and for $D_{A,B}=0$ they are
\begin{align}
\Psi_{A}\lr{r,\theta} &=
\frac{rV_c}{C} \left\{
a_1 (\theta - \pi) \sin\lr{ \theta-\Phi} +
a_2 \lr{\theta - \Phi} \sin\theta  \right\},\\
\Psi_{B}\lr{r,\theta} &=
\frac{rV_c}{C} \left\{
b_1 \theta \sin( \theta - \Phi ) +
b_2 \lr{ \theta - \Phi }  \sin\theta \right\},
\end{align}
where
\begin{align}
C &= \Phi\lr{ \pi - \Phi } - \sin^2 \Phi + \lr{ 2\Phi - \pi } \cos \Phi \sin \Phi,\\
a_1 &= \Phi  \cos\Phi - \sin \Phi,\\
a_2 &= \cos \Phi \sin \Phi  - \Phi,\\
b_1 &= \lr{\Phi-\pi}  \cos\Phi - \sin \Phi,\\
b_2 &= \cos \Phi \sin \Phi + \pi - \Phi.
\end{align}
We note that the specific forms of these solutions in Ref.~\onlinecite{seppecher} have minor typographical errors (separate from the discussions in Sec.~\ref{sec:seppecher}), so will not agree directly with our solutions above.

% 
% \bibliographystyle{DNS_unsrt_abbrv.bst}
% \bibliography{pof_CHbib.bib}

\begin{thebibliography}{10}

\bibitem{Dussan79}
E.~B. Dussan~V.
\newblock On the spreading of liquids on solid surfaces: Static and dynamic
  contact lines.
\newblock {\em Annu. Rev. Fluid Mech.}, 11(1):371--400, 1979.

\bibitem{deGennesrev}
P.~G. de~Gennes.
\newblock Wetting: statics and dynamics.
\newblock {\em Rev. Mod. Phys.}, 57:827--863, 1985.

\bibitem{blake2006physics}
T.~Blake.
\newblock The physics of moving wetting lines.
\newblock {\em J. Colloid Interface Sci.}, 299(1):1--13, 2006.

\bibitem{BonnEggers}
D.~Bonn, J.~Eggers, J.~Indekeu, J.~Meunier, and E.~Rolley.
\newblock Wetting and spreading.
\newblock {\em Rev. Mod. Phys.}, 81:739--805, 2009.

\bibitem{DussanDavis}
E.~B. Dussan~V. and S.~H. Davis.
\newblock On the motion of a fluid-fluid interface along a solid surface.
\newblock {\em J. Fluid Mech.}, 65(1):71--95, 1974.

\bibitem{ShikhSingualarities06}
Y.~D. Shikhmurzaev.
\newblock Singularities at the moving contact line. {M}athematical, physical
  and computational aspects.
\newblock {\em Physica D}, 217(2):121--133, 2006.

\bibitem{HuhScriv71}
C.~Huh and L.~E. Scriven.
\newblock Hydrodynamic model of steady movement of a solid / liquid / fluid
  contact line.
\newblock {\em J. Colloid Interface Sci.}, 35(1):85--101, 1971.

\bibitem{Navier}
C.-L. Navier.
\newblock M\'{e}moire sur les lois du mouvement des fluides.
\newblock {\em Mem. Acad. Sci. Inst. Fr.}, 6:389--440, 1823.

\bibitem{SavvaPrecursorSlip}
N.~Savva and S.~Kalliadasis.
\newblock Dynamics of moving contact lines: A comparison between slip and
  precursor film models.
\newblock {\em Europhys. Lett.}, 94(6):64004, 2011.

\bibitem{anderson_rev}
D.~M. Anderson, G.~B. McFadden, and A.~A. Wheeler.
\newblock Diffuse-interface methods in fluid mechanics.
\newblock {\em Annu. Rev. Fluid Mech.}, 30:139--165, 1998.

\bibitem{seppecher}
P.~Seppecher.
\newblock Moving contact lines in the {C}ahn-{H}illiard theory.
\newblock {\em Int. J. Eng. Sci.}, 34(9):977--992, 1996.

\bibitem{jacqmin}
D.~Jacqmin.
\newblock Contact-line dynamics of a diffuse fluid interface.
\newblock {\em J. Fluid Mech.}, 402:57--88, 2000.

\bibitem{BriantYeomansEarly}
A.~Briant.
\newblock {L}attice {B}oltzmann simulations of contact line motion in a
  liquid-gas system.
\newblock {\em Philos. T. Roy. Soc. A}, 360(1792):485--495, 2002.

\bibitem{BriantYeomans1}
A.~J. Briant, A.~J. Wagner, and J.~M. Yeomans.
\newblock {L}attice {B}oltzmann simulations of contact line motion. {I}.
  {L}iquid-gas systems.
\newblock {\em Phys. Rev. E}, 69:031602, 2004.

\bibitem{jasnowvinals}
D.~Jasnow and J.~Vi{\~{n}}als.
\newblock Coarse-grained description of thermo-capillary flow.
\newblock {\em Phys. Fluids}, 8(3):660--669, 1996.

\bibitem{KhatavkarJFM}
V.~V. Khatavkar, P.~D. Anderson, and H.~E.~H. Meijer.
\newblock Capillary spreading of a droplet in the partially wetting regime
  using a diffuse-interface model.
\newblock {\em J. Fluid Mech.}, 572:367--387, 2007.

\bibitem{DingSpeltJFM}
H.~Ding and P.~D.~M. Spelt.
\newblock Inertial effects in droplet spreading: a comparison between
  diffuse-interface and level-set simulations.
\newblock {\em J. Fluid Mech.}, 576:287--296, 2007.

\bibitem{YueZhouFeng}
P.~Yue, C.~Zhou, and J.~J. Feng.
\newblock Sharp-interface limit of the {C}ahn--{H}illiard model for moving
  contact lines.
\newblock {\em J. Fluid Mech.}, 645:279--294, 2010.

\bibitem{EvansReview}
R.~Evans.
\newblock The nature of the liquid-vapour interface and other topics in the
  statistical mechanics of non-uniform, classical fluids.
\newblock {\em Adv. Phys.}, 28(2):143--200, 1979.

\bibitem{HendersonBook}
D.~Henderson.
\newblock {\em {Fundamentals of Inhomogeneous Fluids}}.
\newblock Dekker, New York, 1st. edition, 1992.

\bibitem{Andreas1}
A.~Nold, A.~Malijevsk\'y, and S.~Kalliadasis.
\newblock Wetting on a spherical wall: Influence of liquid-gas interfacial
  properties.
\newblock {\em Phys. Rev. E}, 84:021603, 2011.

\bibitem{yatsyshin1}
P.~Yatsyshin, N.~Savva, and S.~Kalliadasis.
\newblock Spectral methods for the equations of classical density-functional
  theory: Relaxation dynamics of microscopic films.
\newblock {\em J. Chem. Phys.}, 136(12):124113, 2012.

\bibitem{AlikakosBatesChen}
N.~D. Alikakos, P.~W. Bates, and X.~Chen.
\newblock Convergence of the {C}ahn--{H}illiard equation to the {H}ele--{S}haw
  model.
\newblock {\em Arch. Rational Mech. Anal.}, 128(2):165--205, 1994.

\bibitem{YueFengPoF}
P.~Yue and J.~J. Feng.
\newblock Wall energy relaxation in the {C}ahn--{H}illiard model for moving
  contact lines.
\newblock {\em Phys. Fluids}, 23(1):012106, 2011.

\bibitem{christophe_marc}
C.~Wylock, M.~Pradas, B.~Haut, P.~Colinet, and S.~Kalliadasis.
\newblock Disorder-induced hysteresis and nonlocality of contact line motion in
  chemically heterogeneous microchannels.
\newblock {\em Phys. Fluids}, 24(3):032108, 2012.

\bibitem{epj_entire}
M.~G. {Velarde}.
\newblock {Discussion and Debate: Wetting and Spreading Science - quo vadis?
  [Special issue]}.
\newblock {\em Eur. Phys. J. Special Topics}, 197:1--2, 2011.

\bibitem{ShikhBook}
Y.~D. Shikhmurzaev.
\newblock {\em {Capillary Flows with Forming Interfaces}}.
\newblock Taylor \& Francis, London, 2008.

\bibitem{epj1}
Y.~D. Shikhmurzaev.
\newblock Some dry facts about dynamic wetting.
\newblock {\em Eur. Phys. J. Special Topics}, 197:47--60, 2011.

\bibitem{epj5}
Y.~D. Shikhmurzaev.
\newblock Discussion notes.
\newblock {\em Eur. Phys. J. Special Topics}, 197:73--74, 2011.

\bibitem{epj6}
Y.~D. Shikhmurzaev.
\newblock {Discussion Notes on ``Some singular errors near the contact line
  singularity, and ways to resolve both'', by L.M. Pismen}.
\newblock {\em Eur. Phys. J. Special Topics}, 197:75--80, 2011.

\bibitem{epj8}
Y.~D. Shikhmurzaev.
\newblock {Discussion Notes: On capillarity and slightly beyond}.
\newblock {\em Eur. Phys. J. Special Topics}, 197:85--87, 2011.

\bibitem{epj11}
Y.~D. Shikhmurzaev.
\newblock {Discussion Notes on ``Note on thin film equations for solutions and
  suspensions'', by U. Thiele}.
\newblock {\em Eur. Phys. J. Special Topics}, 197:221--225, 2011.

\bibitem{epj4}
U.~Thiele.
\newblock {Discussion Notes: Thoughts on mesoscopic continuum models}.
\newblock {\em Eur. Phys. J. Special Topics}, 197:67--71, 2011.

\bibitem{epjpismen}
L.~M. Pismen.
\newblock Some singular errors near the contact line singularity, and ways to
  resolve both.
\newblock {\em Eur. Phys. J. Special Topics}, 197:33--36, 2011.

\bibitem{epj3}
L.~M. Pismen.
\newblock {Discussion Notes on ``Some dry facts about dynamic wetting'', by
  Y.D. Shikhmurzaev}.
\newblock {\em Eur. Phys. J. Special Topics}, 197:63--65, 2011.

\bibitem{epj7}
Y.~Pomeau.
\newblock {Discussion Notes: More (and last remarks) on the debate on
  capillarity}.
\newblock {\em Eur. Phys. J. Special Topics}, 197:81--83, 2011.

\bibitem{pomeauphasefield}
Y.~Pomeau.
\newblock Discussion notes: Phase field models and moving contact line in the
  long perspective.
\newblock {\em Eur. Phys. J. Special Topics}, 197(1):11--13, 2011.

\bibitem{epj12}
T.~D. Blake.
\newblock Discussion notes: A more collaborative approach to the moving
  contact-line problem?
\newblock {\em Eur. Phys. J. Special Topics}, 197:343--345, 2011.

\bibitem{PismenPomeau}
L.~M. Pismen and Y.~Pomeau.
\newblock Disjoining potential and spreading of thin liquid layers in the
  diffuse-interface model coupled to hydrodynamics.
\newblock {\em Phys. Rev. E}, 62:2480--2492, 2000.

\bibitem{myEPJE}
D.~N. Sibley, A.~Nold, N.~Savva, and S.~Kalliadasis.
\newblock On the moving contact line singularity: Asymptotics of a
  diffuse-interface model.
\newblock {\em Eur. Phys. J. E}, 36:26, 2013.

\bibitem{cahn77}
J.~W. Cahn.
\newblock Critical point wetting.
\newblock {\em J. Chem. Phys.}, 66(8):3667--3672, 1977.

\bibitem{Pismenmeso}
L.~Pismen.
\newblock Mesoscopic hydrodynamics of contact line motion.
\newblock {\em Colloid. Surface. A}, 206(1–3):11 -- 30, 2002.

\bibitem{antoniojfm}
A.~Pereira and S.~Kalliadasis.
\newblock Equilibrium gas–-liquid–-solid contact angle from
  density-functional theory.
\newblock {\em J. Fluid Mech.}, 692:53--77, 2012.

\bibitem{Noether}
E.~Noether.
\newblock Invariante variationsprobleme.
\newblock {\em Nachr. v. d. Ges. d. Wiss. zu G\"{o}ttingen}, pages 235--257,
  1918.

\bibitem{Markus}
M.~Schmuck, M.~Pradas, G.~A. Pavliotis, and S.~Kalliadasis.
\newblock Upscaled phase-field models for interfacial dynamics in strongly
  heterogeneous domains.
\newblock {\em Proc. R. Soc. A}, 468(2147):3705--3724, 2012.

\bibitem{YueZhouFeng04}
P.~Yue, J.~J. Feng, C.~Liu, and J.~Shen.
\newblock A diffuse-interface method for simulating two-phase flows of complex
  fluids.
\newblock {\em J. Fluid Mech.}, 515:293--317, 2004.

\bibitem{qianliqgas}
X.~Xu and T.~Qian.
\newblock Contact line motion in confined liquid--gas systems: Slip versus
  phase transition.
\newblock {\em J. Chem. Phys.}, 133(20):204704, 2010.

\bibitem{CH58}
J.~W. Cahn and J.~E. Hilliard.
\newblock Free energy of a nonuniform system. {I}. {I}nterfacial free energy.
\newblock {\em J. Chem. Phys.}, 28(2):258--267, 1958.

\bibitem{QianWangShengJFM}
T.~Qian, X.-P. Wang, and P.~Sheng.
\newblock A variational approach to moving contact line hydrodynamics.
\newblock {\em J. Fluid Mech.}, 564:333--360, 2006.

\bibitem{QianWangShengGNBCfirst}
T.~Qian, X.-P. Wang, and P.~Sheng.
\newblock Molecular scale contact line hydrodynamics of immiscible flows.
\newblock {\em Phys. Rev. E}, 68:016306, 2003.

\bibitem{KassnerPaperSI}
C.~Gugenberger, R.~Spatschek, and K.~Kassner.
\newblock Comparison of phase-field models for surface diffusion.
\newblock {\em Phys. Rev. E}, 78:016703, 2008.

\bibitem{batchelor}
G.~K. Batchelor.
\newblock {\em An Introduction to Fluid Dynamics}.
\newblock Cambridge University Press, Cambridge, 2000.

\bibitem{fredholm}
I.~Fredholm.
\newblock Sur une classe d'\'{e}quations fonctionnelles.
\newblock {\em Acta Mathematica}, 27:365--390, 1903.

\bibitem{Moffatt}
H.~K. Moffatt.
\newblock Viscous and resistive eddies near a sharp corner.
\newblock {\em J. Fluid Mech.}, 18(01):1--18, 1964.

\bibitem{YueFengEPJ}
P.~Yue and J.~Feng.
\newblock Can diffuse-interface models quantitatively describe moving contact
  lines?
\newblock {\em Eur. Phys. J. Special Topics}, 197:37--46, 2011.

\bibitem{My_Shikh}
D.~N. Sibley, N.~Savva, and S.~Kalliadasis.
\newblock Slip or not slip? a methodical examination of the interface formation
  model using two-dimensional droplet spreading on a horizontal planar
  substrate as a prototype system.
\newblock {\em Phys. Fluids}, 24(8):082105, 2012.

\bibitem{Savva09}
N.~Savva and S.~Kalliadasis.
\newblock Two-dimensional droplet spreading over topographical substrates.
\newblock {\em Phys. Fluids}, 21(9):092102, 2009.

\bibitem{Savva10}
N.~Savva, S.~Kalliadasis, and G.~A. Pavliotis.
\newblock Two-dimensional droplet spreading over random topographical
  substrates.
\newblock {\em Phys. Rev. Lett.}, 104:084501, 2010.

\bibitem{SavvaPavliotisKalliadasis1}
N.~Savva, G.~A. Pavliotis, and S.~Kalliadasis.
\newblock Contact lines over random topographical substrates. {P}art 1.
  {S}tatics.
\newblock {\em J. Fluid Mech.}, 672:358--383, 2011.

\bibitem{SavvaPavliotisKalliadasis2}
N.~Savva, G.~A. Pavliotis, and S.~Kalliadasis.
\newblock Contact lines over random topographical substrates. {P}art 2.
  {D}ynamics.
\newblock {\em J. Fluid Mech.}, 672:384--410, 2011.

\bibitem{RajChemHet}
R.~Vellingiri, N.~Savva, and S.~Kalliadasis.
\newblock Droplet spreading on chemically heterogeneous substrates.
\newblock {\em Phys. Rev. E}, 84:036305, 2011.

\bibitem{Herde12}
D.~Herde, U.~Thiele, S.~Herminghaus, and M.~Brinkmann.
\newblock Driven large contact angle droplets on chemically heterogeneous
  substrates.
\newblock {\em Europhys. Lett.}, 100:16002, 2012.

\bibitem{DupuisYeomans}
A.~Dupuis and J.~M. Yeomans.
\newblock Modeling droplets on superhydrophobic surfaces: Equilibrium states
  and transitions.
\newblock {\em Langmuir}, 21(6):2624--2629, 2005.

\bibitem{KusumaatmajaYeomans}
H.~Kusumaatmaja, M.~L. Blow, A.~Dupuis, and J.~M. Yeomans.
\newblock The collapse transition on superhydrophobic surfaces.
\newblock {\em Europhys. Lett.}, 81(3):36003, 2008.

\bibitem{KhatavkarChemEngSci}
V.~V. Khatavkar, P.~D. Anderson, and H.~E.~H. Meijer.
\newblock On scaling of diffuse-interface models.
\newblock {\em Chem. Eng. Sci.}, 61(8):2364 -- 2378, 2006.

\bibitem{AbelsGarckeGrun12}
H.~Abels, H.~Garcke, and G.~Gr{\"u}n.
\newblock Thermodynamically consistent, frame indifferent diffuse interface
  models for incompressible two-phase flows with different densities.
\newblock {\em Math. Mod. Meth. Appl. Sci.}, 22(03):1150013, 2012.

\bibitem{Magaletti_etal_2013}
F.~Magaletti, F.~Picano, M.~Chinappi, L.~Marino, and C.~M. Casciola.
\newblock The sharp-interface limit of the
  {C}ahn--{H}illiard/{N}avier--{S}tokes model for binary fluids.
\newblock {\em J. Fluid Mech.}, 714:95--126, 2013.

\bibitem{WangWangSharp}
X.~Wang and Y.~Wang.
\newblock The sharp interface limit of a phase field model for moving contact
  line problem.
\newblock {\em Methods Appl. Anal.}, 14:287--294, 2007.

\bibitem{WeidnerSchwartz}
D.~E. Weidner and L.~W. Schwartz.
\newblock Contact-line motion of shear-thinning liquids.
\newblock {\em Phys. Fluids}, 6:3535--3538, 1994.

\bibitem{YueetalGooch}
P.~Yue, C.~Zhou, J.~J. Feng, C.~F. Ollivier-Gooch, and H.~H. Hu.
\newblock Phase-field simulations of interfacial dynamics in viscoelastic
  fluids using finite elements with adaptive meshing.
\newblock {\em J. Comput. Phys.}, 219(1):47--67, 2006.

\bibitem{RexLowenEPJE}
M.~Rex and H.~L{\"o}wen.
\newblock Dynamical density functional theory for colloidal dispersions
  including hydrodynamic interactions.
\newblock {\em Eur. Phys. J. E}, 28:139--146, 2009.

\bibitem{BenPRL}
B.~D. Goddard, A.~Nold, N.~Savva, G.~A. Pavliotis, and S.~Kalliadasis.
\newblock General dynamical density functional theory for classical fluids.
\newblock {\em Phys. Rev. Lett.}, 109:120603, 2012.

\bibitem{BenJPCM}
B.~D. Goddard, A.~Nold, N.~Savva, P.~Yatsyshin, and S.~Kalliadasis.
\newblock Unification of dynamic density functional theory for colloidal fluids
  to include inertia and hydrodynamic interactions: derivation and numerical
  experiments.
\newblock {\em J. Phys.: Condens. Matter}, 25(3):035101, 2013.

\bibitem{PetrPRE}
P.~Yatsyshin, N.~Savva, and S.~Kalliadasis.
\newblock Geometry-induced phase transition in fluids: Capillary prewetting.
\newblock {\em Phys. Rev. E}, 87:020402(R), 2013.

\bibitem{barker_henderson}
J.~A. Barker and D.~Henderson.
\newblock Perturbation theory and equation of state for fluids. {II}. {A}
  successful theory of liquids.
\newblock {\em J. Chem. Phys.}, 47(11):4714--4721, 1967.

\bibitem{AndersonSI}
D.~Anderson, G.~McFadden, and A.~Wheeler.
\newblock A phase-field model with convection: sharp-interface asymptotics.
\newblock {\em Physica D}, 151(2–4):305 -- 331, 2001.

\bibitem{Pego89}
R.~L. Pego.
\newblock Front migration in the nonlinear {C}ahn--{H}illiard equation.
\newblock {\em Proc. R. Soc. Lond. A}, 422(1863):261--278, 1989.

\bibitem{CaginalpFife}
G.~Caginalp and P.~C. Fife.
\newblock Dynamics of layered interfaces arising from phase boundaries.
\newblock {\em SIAM J. Appl. Math.}, 48(3):506--518, 1988.

\bibitem{Lowengrub98}
J.~Lowengrub and L.~Truskinovsky.
\newblock Quasi-incompressible {C}ahn--{H}illiard fluids and topological
  transitions.
\newblock {\em Proc. R. Soc. Lond. A}, 454(1978):2617--2654, 1998.

\bibitem{Folchsharp}
R.~Folch, J.~Casademunt, A.~Hern\'andez-Machado, and L.~Ram\'{i}rez-Piscina.
\newblock Phase-field model for hele-shaw flows with arbitrary viscosity
  contrast. {I}. {T}heoretical approach.
\newblock {\em Phys. Rev. E}, 60:1724--1733, 1999.

\end{thebibliography}

\end{document}